\documentclass[12pt]{article}
\newlength{\abstractwidth}
\renewcommand{\thefootnote}{\fnsymbol{footnote}}
\renewcommand{\thanks}[1]{\footnote{#1}}
\newcommand{\starttext}{
\setcounter{footnote}{0}
\renewcommand{\thefootnote}{\arabic{footnote}}}

\newcommand{\bea}{\begin{eqnarray}}
\newcommand{\eea}{\end{eqnarray}}
\newcommand{\ee}{\end{equation}}
\newcommand{\be}{\begin{equation}}

\newcommand{\sm}{\smallskip}

\def\cB{{\cal B}}

\def\cG{{\cal G}}
\def\cH{{\cal H}}

\def\cN{{\cal N}}

\def\cS{{\cal S}}

\def\bZ{{\bf Z}}

\def\Im{{\rm Im}}

\def\det{{\rm det}}

\def\half{ {1\over 2}}
\def\p{\partial}
\def\l({\left(}
\def\r){\right)}

\def\a{\alpha}
\def\b{\beta}

\def\ep{\varepsilon}

\def\g{\gamma}
\def\o{\omega}

\def\G{\Gamma}

\def\g{\gamma}
\def\s{\sigma}

\def\t{\tau}

\def\ch{{\rm ch}}

\def\th{{\rm th}}

\def\no{\nonumber}

\begin{document}
\starttext
\setcounter{footnote}{0}

\begin{flushright}
26 January  2010
\end{flushright}

\bigskip

\begin{center}

{\Large \bf Rigidity of $SU(2,2|2)$-symmetric solutions in Type IIB \footnote{This
work was supported in part by National Science Foundation grant PHY-07-57702.}}

\medskip

\vskip .5in

{\large \bf Eric D'Hoker and  Yu Guo}

\vskip .15in

{ \sl Department of Physics and Astronomy }\\
{\sl University of California, Los Angeles, CA 90095, USA}\\
{\tt \small dhoker@physics.ucla.edu; guoyu@physics.ucla.edu}

\end{center}

\vskip .4in

\begin{abstract}

\vskip 0.1in

We investigate the existence of half-BPS solutions in Type IIB supergravity which are
invariant under the superalgebra $SU(2,2|2)$ realized on either
$AdS_5 \times S^2 \times S^1$ or $AdS_5 \times S^3$ warped over a
Riemann surface $\Sigma$ with boundary. We prove that, in both cases,
the only solution is $AdS_5 \times S^5$ itself. We argue that this result
provides evidence for the non-existence of fully back-reacted intersecting
D3/D7 branes with either $AdS_5 \times S^2 \times S^1\times \Sigma $ or
$AdS_5 \times S^3 \times \Sigma $ near-horizon limits.

\end{abstract}

\newpage


\newpage

\section{Introduction}
\setcounter{equation}{0}

The AdS/CFT correspondence \cite{Maldacena:1997re,Gubser:1998bc,Witten:1998qj}
(for reviews, see \cite{Aharony:1999ti,D'Hoker:2002aw}) allows one to study gauge
theories, in the large $N$ limit and at strong `t Hooft coupling, in terms of dual supergravity solutions in Type IIB or M-theory. Superconformal Yang-Mills theories are of special
interest. They are dual to supergravity solutions whose space-time exhibits a (warped)
$AdS$ factor, and which are invariant under a conformal super algebra. This case
encompasses some of the best understood of all the AdS/CFT dualities.

\sm

Substantial progress has been made in recent years in obtaining  half-BPS supergravity
solutions which are dual to Yang-Mills theories in various dimensions with 16
conformal supersymmetries.  Here, the constraints imposed by supersymmetry are strong enough to allow for large classes of exact solutions, while the dynamics remains 
sufficiently rich
to allow for large moduli spaces. A systematic construction, in both  Type IIB and
M-theory, was initiated in \cite{Lin:2004nb}, 
where solutions dual to local half-BPS
operator insertions in $\cN=4$ super-Yang-Mills were obtained, as were
$AdS_5$ solutions in M-theory which are dual to $\cN=2$ superconformal
Yang-Mills theories in 4-dimensions  (see also \cite{Gauntlett:2005ww}). 
Subsequently, general classes of exact half-BPS
solutions dual to planar interfaces
\cite{Gomis:2006cu,Lunin:2007ab,D'Hoker:2007xy,D'Hoker:2007xz,D'Hoker:2008wc,
D'Hoker:2008qm,D'Hoker:2009gg},
Wilson loops \cite{Yamaguchi:2006te, Lunin:2006xr, D'Hoker:2007fq},
and surface operators \cite{Buchbinder:2007ar,D'Hoker:2008qm}
were constructed  in Type IIB and in M-theory.  These solutions form families, and
exhibit a wealth of topological and metrical structure,
and are characterized by interesting, and generally complicated,
moduli spaces. Generalizations of Janus solutions and their CFT duals to include
the $\theta$-angle, as well as a systematic study of their S-duality properties
may be found in \cite{Gaiotto:2008sa,Gaiotto:2008sd,Gaiotto:2008ak}.

\sm

A general correspondence, in  supergravities with 32 supersymmetries, 
between half-BPS solutions and certain Lie superalgebras with 16 fermionic generators
was proposed in \cite{D'Hoker:2008ix} for both Type IIB and M-theory.
For the Type IIB case of interest here, semi-simple Lie superalgebras
$\cH$ with 16 fermionic generators which are subalgebras of $PSU(2,2|4)$ are
related to families of half-BPS solutions to Type IIB supergravity which are invariant
under $\cH$, and are locally asymptotic to $AdS_5 \times S^5$. Given that  there
are a finite number of such subalgebras $\cH$ (they were completely catalogued in
\cite{D'Hoker:2008ix}, and references therein), a precise classification of such half-BPS
solutions becomes available. (Earlier classifications in terms of Killing spinors,
their bilinears, and $G$-structures may be found in \cite{Gauntlett:2002sc,Gauntlett:2002fz,Gauntlett:2004zh,Gauntlett:2005bn}.) The special classes of exact solutions discussed
in the preceding paragraph all fit into this scheme. The superalgebra classification
predicts, however, cases where a suitable subalgebra $\cH$ exists, but no corresponding
supergravity solutions have been identified yet.

\sm

Another approach to the classification of half-BPS solutions is in terms of the near-horizon
limit of various brane intersections, in terms of which many of the known half-BPS solutions
may be understood. The brane picture offers a powerful method for understanding the global and qualitative properties of the supergravity solutions, and has led recently to a unified picture of
M-theory solutions dual to $\cN=2$ superconformal Yang-Mills theories in 4 dimensions \cite{Gaiotto:2009gz,Alday:2009fs,Gaiotto:2009fs}. All known exact half-BPS solutions have known brane realizations.
The brane picture suggests cases, however, of half-BPS near horizon limits of
brane intersection where no corresponding supergravity solutions have been
found yet.

\sm

Perhaps the most important case where the superalgebra and brane pictures
appear to be at odds with the absence of any known solutions is for Type IIB supergravity
solutions which are half-BPS, locally asymptotic to $AdS_5 \times S^5$, and
invariant under  $SU(2,2|2)$. The duals to such solutions would be 4-dimensional
$\cN=2$ super-Yang-Mills theories with exact conformal invariance. Their brane
picture would be in terms of the near-horizon limit of stacks of  D3- and D7-branes,
intersecting along a 4-dimensional space-time, or equivalently  in terms of probe D7-branes
inside $AdS_5$.

\sm

The conflict is as follows. The existence of the subalgebra
$SU(2,2|2) \subset PSU(2,2|4)$ clearly makes it possible to have asymptotically
$AdS_5 \times S^5$ half-BPS solutions with this symmetry. The existence and
stability of probe D7-branes in $AdS_5$ with $SU(2,2|2)\times SU(2)$ symmetry
also supports the possibility of having such solutions, though at the expense
of including an extra bosonic $SU(2)$ symmetry group not mandated
by superconformal invariance. The question is whether such probe brane solutions
extend to fully back-reacted
supergravity solutions where the D7-branes dissolve into regular flux solutions.

\sm

Several arguments have been presented
\cite{Kirsch:2005uy,Buchbinder:2007ar,Harvey:2008zz}
(following earlier work \cite{Aharony:1998xz,Grana:2001xn})
that no such fully back-reacted solutions corresponding to D7-branes should exist.
A first argument against is that D7-branes produce flavor multiplets
\cite{Karch:2002sh,Constable:2002xt}
transforming under the fundamental representation of the gauge group in the dual gauge
theory. Their contribution to the renormalization group $\beta$-function for the gauge
coupling is non-vanishing, and thus breaks the exact conformal invariance needed
to maintain the $AdS_5$ factor on the gravity side.
A second argument \cite{Sheikh:unpub} makes use of local 1/2 BPS-operators:
the invariance superalgebra $\mathcal{H}$ for half-BPS solutions which are genuinely
asymptotic to $AdS_5\times S^5$  must have rank strictly less than the rank 6 of
$PSU(2, 2|4)$. This argument leads us to conclude that solutions with
$SU(2,2|2)\times SU(2)$ symmetry cannot be asymptotic to $AdS_5 \times S^5$,
though there appears to be no obstruction to the
existence of {\sl locally asymptotic} solutions.

\sm

In the present paper, we shall focus on half-BPS Type IIB supergravity solutions with
bosonic symmetries  $SO(2,4) \times SO(3) \times SO(2)$ and
$SO(2,4) \times SO(4)\times SO(2) $ respectively corresponding to the maximal bosonic symmetries of $SU(2,2|2)$ and $SU(2,2|2)\times SU(2)$.
These cases emerge from the general classification of
half-BPS solutions via superalgebra arguments \cite{D'Hoker:2008ix}.
Following the general construction of the space-time Ans\"atze given there,
the symmetries $SO(2,4) \times SO(3) \times SO(2)$ and
$SO(2,4) \times SO(4)\times SO(2) $ are realized as isometries of the spaces
$AdS_5 \times S^2 \times S^1$ and $AdS_5 \times S^3 \times S^1$ respectively.

\sm

We shall prove here that the only solution with at least 16 supersymmetries for
space-times of the form $AdS_5 \times S^2 \times S^1\times \Sigma$ and
$AdS_5 \times S^3 \times \Sigma$, invariant respectively under
$SO(2,4)\times SO(3)\times SO(2) $ and $SO(2,4) \times SO(4)$,
is just the maximally supersymmetric solution $AdS_5 \times S^5$. Here,
the products $AdS_5 \times S^2 \times S^1$ and $AdS_5 \times S^3$
are warped over the Riemann surface $\Sigma$. It then follows that no
near-horizon limit can exist (in the supergravity limit) for fully back-reacted
intersection solutions of  D3- and D7-branes with those symmetries and
space-time geometries. It is in this sense that the $SU(2,2|2)$-invariant
solutions are {\sl rigid}. Note that rigidity of half-BPS solutions has already
been encountered in analyzing analogous problems in M-theory \cite{D'Hoker:2009my}.

\sm

Finally, we note that there is one possible interesting near-horizon limit of
D3- and D7-brane fully back-reacted solutions that is not yet excluded by
our study. One could imagine a space-time of the form
$AdS_5 \times \left ( S^2 \times_k S^1 \right ) \times \Sigma$
where the product $S^2 \times_k S^1$ is twisted by the Hopf fibration,
with $k \in \bZ$ being the monopole number. For $k=0$, we recover the case
$AdS_5 \times S^2 \times S^1 \times \Sigma$ studied here. For $k \not=0$
one recovers $S^3/Z_k$ topologically, but the metric may be deformed
to that of a squashed $S^3$. For $k=1$, and no squashing, we recover the
case $AdS_5 \times S^3 \times \Sigma$ studied here. In general, however,
the squashing parameter (the ratio of the radius of $S^1$
to that of $S^2$) may be non-trivial, and actually be a function of $\Sigma $.
Such cases with non-trivial fibration and squashing require a generalization of the
construction methods proposed in \cite{D'Hoker:2008ix}, and are clearly
of physical interest. They are considerably more complicated than the cases
treated here; their study will be postponed to future work.

\subsection{Organization}

The remainder of this paper is organized as follows.
In section 2, Type IIB supergravity is briefly reviewed, mostly to fix notations,
and the $AdS_5 \times S^2 \times S^1\times \Sigma $ Ansatz is implemented on
the Type IIB supergravity fields.
In section 3, the BPS equations are reduced on this Ansatz; they are the
starting point of the exact solutions. In section 4, we show that the vanishing 
of the 3-form field strength implies the maximally supersymmetric solutions 
$AdS_5 \times  S^5$. In section 5, we solve for the metric factors of the 
$AdS_5 \times S^2\times S^1$ spaces, and express them in terms of
bilinears in the supersymmetry spinors.
In section 6, the vanishing of several types of Hermitian forms in the 
supersymmetry spinors are derived from the reduced BPS equations.
In section 7, with the help of these vanishing Hermitian forms, it is shown that
the only viable solution for the $AdS_5 \times S^2  \times S^1\times \Sigma$ Ansatz
is $AdS_5 \times S^5$. In section 8, the analysis is extended to the
$AdS_5 \times S^3 \times \Sigma $ Ansatz. It is shown that also here the
only viable half-BPS solution is $AdS_5 \times S^5$.
In Appendix A, we give Clifford algebra representations adapted to the
two different Ans\"atze. In Appendices B and C, we provide the details of reducing BPS
equations for the two different Ans\"atze. In Appendix D and E, technical details
of solving metric factors and vanishing Hermitian forms are presented.

\bigskip

\noindent
{\large \bf Acknowledgments}

\medskip

We wish to thank John Estes, Michael Gutperle, Per Kraus, Darya Krym, David Mateos,
Diego Trancanelli for helpful discussions, and Andrew Royston and M.M. Sheikh Jabbari
for useful correspondence.

\newpage

\section{Ansatz with $SO(2,4) \times SO(3) \times SO(2)$ symmetry}
\setcounter{equation}{0}
\label{two}

In this section, we briefly review Type IIB supergravity, and construct the
Ansatz for supergravity solutions with $SO(2,4) \times SO(3) \times SO(2)$
symmetry.

\subsection{Type IIB supergravity}

The canonical fields of Type IIB supergravity are the metric $g_{MN}$,
and the 4-form $C_{(4)}$, both of which are real fields, the axion/dilaton
scalar field $B$, and the 2-form $B_{(2)}$, both of which are complex fields,
as well as the dilatino $\lambda$ and gravitino  $\psi_M$ spinor fields.
Our  conventions are those of \cite{D'Hoker:2006uu,Schwarz:1983qr}
(see also \cite{Howe:1983sr}). It is convenient to formulate
Type IIB supergravity in terms of the field strengths, Bianchi identities,
and field equations. For the axion/dilaton, we introduce the 1-form fields
$P,Q$,
\bea
\label{sugra1}
P  & = & f_B^2 d B \hskip 0.45in  \hskip 1in
f_B^2=(1-|B|^2)^{-1}
\no \\
Q  & = & f_B^2 \Im( B d  \bar B)
\eea
while for the 2- and 4-form fields we introduce the field strengths,
\bea
\label{GF5}
G & = &  f_B(F_{(3)} - B \bar F_{(3)} ) \hskip 1in F_{(3)} = d B_{(2)}
\no \\
F_{(5)} & = & dC_{(4)} + { i \over 16} \left ( B_{(2)} \wedge \bar
F_{(3)} - \bar B_{(2)} \wedge  F_{(3)} \right )
\label{F_5}
\eea
The scalar field $B$ is related to the complex string coupling $\tau$,
the axion $\chi$, and dilaton $\Phi$.
In terms of the composite fields $P,Q$, and $G$, there
are Bianchi identities given as follows,
\bea
0 &=& dP-2i Q\wedge P
\label{bianchi1} \\
0 &=& d Q + i P\wedge \bar P
\label{bianchi2} \\
0 &=& d G - i Q\wedge G +  P\wedge \bar G
\label{bianchi3} \\
0 &=& d F_{(5)} -  {i\over 8} G \wedge \bar G
\label{bianchi4}
\eea
The field strength $F_{(5)}$ is required to be self-dual,
\bea
\label{SDeq}
F_{(5) } = * F_{(5) }
\eea
The field equations are given by,
\bea
0 & = & \nabla ^M P_M -
2i Q^M P_M + {1\over 24} G_{MNP }G^{MNP}
\label{Peq}
\\
0 & = & \nabla ^P  G_{MNP } -i Q^P G_{MNP} - P^P \bar G_{MNP } +
{2\over 3} i F_{(5)MNPQR }G^{PQR}
\label{Geq}
\\
0 & = & R_{MN } - P_M  \bar P_N  - \bar P_M  P_N - {1\over 6}
(F_{(5)}^2) _{MN}
\no \\ && \hskip .5in - {1\over 8} (G_M {} ^{PQ }
\bar G_{N PQ } + {\bar G_M} {} ^{ PQ } G_{N PQ }) +{1\over 48 }
g_{MN } G^{PQR } \bar G_{PQR }
\label{Eeq}
\eea
The BPS equations for bosonic solutions are obtained by setting
the supersymmetry variations $\delta \lambda$ and $\delta \psi _M$
of the fermionic fields to zero, in the presence of vanishing fermi
fields,\footnote{Throughout, we shall use the notation $\G \cdot T
\equiv \G^{M_1 \cdots M_p} T_{M_1 \cdots M_p}$ for the contraction
of any antisymmetric tensor field $T$ of rank $p$ and the
$\G$-matrix of the same rank.}
\bea
\label{BPS}
0 &=& i (\G \cdot P) \cB^{-1} \ep^* -{i\over 24} (\G \cdot G) \ep
\no \\
0 &=& D _M  \ep + {i\over 480} \left ( \G \cdot F_{(5)} \right )
\Gamma_M  \ep -{1\over 96} \left ( \Gamma_M (\G \cdot G) + 2 (\G
\cdot G) \G^M \right ) \cB^{-1} \ep^*
\eea
Here, ${\cal B}$ is the complex conjugation matrix of the Clifford
algebra.\footnote{The matrix $\cB$ is defined by $\cB \cB^*=I$ and
$\cB \Gamma ^M \cB^{-1} = (\Gamma ^M)^*$; see Appendix A  for our
$\Gamma$-matrix conventions. Throughout, complex conjugation will be
denoted by {\sl bar} for functions, and by {\sl star} for spinors.}

\sm

Type IIB supergravity is invariant under $SU(1,1) \sim SL(2,{\bf
R})$ symmetry, which leaves $g_{MN}$ and $C_{(4)}$ invariant,
acts by M\"obius transformation on the field $B$, and linearly on
$B_{(2)}$, 
\bea 
\label{stransf} 
B           & \to &  {u B  + v \over \bar v B + \bar u}
\no \\
B_{(2)} & \to &  u B_{(2)} + v \bar B_{(2)}
\eea
with $u,v \in {\bf C}$ and $\bar u u - \bar v v=1$.
The transformation rules for the composite fields are \cite{D'Hoker:2006uu},
\bea
\label{su11a}
P & \to &  e^{2 i \theta} P
\no \\
Q & \to &  Q + d \theta
\no \\
G & \to &  e^{i \theta} G
\eea
where the phase $\theta$ is defined by
\bea
\label{su11b}
e^{2 i \theta} =  {v \bar B + u \over \bar v B + \bar u}
\eea
In this form, the transformation rules clearly exhibit the $U(1)_q$ gauge
transformation that accompanies the  global $SU(1,1)$
transformations.

\subsection{The $SO(2,4) \times SO(3) \times SO(2)$-invariant Ansatz}

We seek the most general Ansatz in Type IIB supergravity with
$ SO(2,4) \times SO(3) \times SO(2)$ symmetry. The $SO(2,4)$-factor
requires the geometry to contain an $AdS_5$-factor, while the $SO(3)$-
and $SO(2)$-factors respectively require  $S^2$- and $S^1$-factors,
all warped over a 2-dimensional surface $\Sigma$. The total space-time
then has the structure,
\bea
AdS_5 \times S^2  \times \Sigma \times S^1
\eea
Following the general constructing as in \cite{D'Hoker:2007xy}, the Ansatz will be given in terms
of $ SO(2,4) \times SO(3) \times SO(2)$ invariant supergravity fields, which are functions
on the surface $\Sigma$. The Ansatz for the metric is given by,
\bea
ds^2 = f_5^2 ds^2 _{AdS_5}
+ f_2 ^2 ds^2 _{S^2} +  ds^2 _\Sigma + f_1 ^2 ds^2_{S^1}
\eea
where $f_5,f_2$ and $f_1$ are real functions on $\Sigma$. Orthonormal frames are defined by,
\bea
\label{frame2}
e^m & = & f_5 \, \hat e^m \hskip 1in m=0,1,2,3,4
\no \\
e^{i} \, & = & f_2 \, \hat e^{i} \hskip 1.1in i =5,6
\no \\
e^a &= &\rho \, \hat e^a \,  \hskip 1.1in a=7,8
\no \\
e^{9} \, & = & f_1 \, \hat e^{9}
\eea
Here, the metrics $ds^2 _{AdS_5}$, $ds^2 _{S^2}$, and $ds^2 _{S^1}$, as
well as the orthonormal frames $\hat e^m$, $\hat e^{i}$, and
$\hat e^9$ refer to the spaces $AdS_5$, $S^2$,
and $S^1$ with unit radius. In particular, we have
\bea
ds^2 _{AdS_5} & = & \eta _{mn} \, \hat e^m \otimes \hat e^n
\no \\
ds^2 _{S^2} ~ & = & \delta _{i j} \, \hat e^{i} \otimes \hat e^{j}
\no \\
ds^2 _\Sigma  ~ & = & \delta _{ab} \, e^{a} \otimes e^{b}
\no \\
ds^2 _{S^1} ~ & = & \hat e^{9} \otimes \hat e^{9}
\eea
where $\eta = {\rm diag}[- + + +]$. Similarly, the dilaton/axion field  $B$ is a function of $\Sigma$ only. As a
result, the 1-forms $P$ and $Q$ have no component along $e^9$,
so that  $p_9 = q_9 = 0$ and their structures is simply given by,
\bea
\label{PQdef}
P& = & p_a e^a
\no \\
Q & = & q_a e^a
\eea
Finally, the most general Ans\"atze for the 3- and 5-form fields $G$ and $F_{(5)}$
consistent with $SO(2,4) \times SO(3) \times SO(2)$ invariance, are given
by,\footnote{Throughout the $SO(2,4) \times SO(3) \times SO(2)$ symmetry case, indices $a,b$ will run over the
values $7,8$, while indices $\tilde a, \tilde b$ will run over the values $7,8,9$.}
\bea
G & = & i g_{\tilde a} e^{56 \tilde a} + h \, e^{789}
=i  g_{a} e^{56a} +  i g_9 e^{569}+ h \, e^{789}
\no \\
F_{(5)} & = & f \left ( e^{01234} - e^{56789} \right )
\eea
Note that the components of $G$ involving any of the directions $0,1,2,3,4$
must vanish by $SO(2,4)$ symmetry, and that the remaining terms
are the only allowed ones by $SO(3)$-invariance.  By
$SO(2,4) \times SO(3) \times SO(2)$-invariance, the coefficient functions
$p_a, q_a, g_a, g_9, h$, and $f$ depend only on $\Sigma$. The functions
$f, q_a$ are real, while $p_a, h, g_a, g_9$ are complex-valued.

\subsection{Immediate constraints from the Bianchi identities}

There are two immediate consequences of the Bianchi identities and field definitions for the
coefficient functions of the Ansatz.

\sm

$\bullet$ The coefficient $g_9$ is determined by the field definitions. This may be
shown directly by observing that the $g_9$-term in $G$ can arise only from
a contribution $b_0 \hat e^{569}$ to $F_{(3)}$ where $b_0$ is a complex function of $\Sigma$.
Closure of $F_{(3)}$ requires $b_0$ to be constant, so that $g_9$ is given by,
\bea
g_9  = -i {f_B \over f_2^2 f_1} \left ( b_0 - B b_0^* \right )
\eea
with $b_0$ an arbitrary complex constant. Note that the corresponding
$B_{(2)}$-field is given by $b_0 \, x^9 \, \hat e^{56}$. This  form is invariant
{\sl  up to a gauge transformation} under the action of $SO(2)$, namely under
a constant shift in the coordinate $x^9$  of the circle $S^1$.

\sm

$ \bullet$ The Bianchi identity for $F_{(5)}$ reduces to $d F_{(5)}=0$,
since $G \wedge \bar G$ vanishes identically on the Ansatz.
Using the closure of $\hat e^{01234}$, we find that $f \, f_5 ^5$ must be constant,
\bea
\label{bianchif}
f \, f_5 ^5 = c_0
\eea
where $c_0$ is an arbitrary real constant.

\section{The reduced BPS equations}
\setcounter{equation}{0}
\label{three}

Half-BPS configurations and solutions in Type IIB supergravity with vanishing
Fermi fields are such that the BPS equations (\ref{BPS})  for the spinor~$\ep$
admit 16 linearly independent solutions.
The spinor $\ep$ is covariant under $SO(2,4) \times SO(3) \times SO(2)$, and
may be built from the Killing spinors on the space $AdS_5 \times S^2 \times S^1$,
where each factor space has unit radius. Once this structure is in place, the reduction
of the full BPS equations of (\ref{BPS}) is standard.

\subsection{Using killing spinors}

We begin by constructing suitable Killing spinors on
$AdS_5 \times S^2$. We define an 8-dimensional spinor representation
$\chi ^{\eta_1, \eta_2}$ of $SO(2,4) \times SO(3)$  by the following equations
\cite{D'Hoker:2007xy,D'Hoker:2008wc},
\bea
\label{KS}
\left ( \hat \nabla _m  - \half \eta_1 \gamma _m \otimes I_2  \right )
\chi ^{\eta _1, \eta _2}
    & = & 0 \hskip 1in m=0,1,2,3,4
\no \\
\left ( \hat \nabla _{i} - {i \over 2} \eta_2 I_4 \otimes \gamma _{i} \right )
\chi ^{\eta _1, \eta _2}
    & = & 0 \hskip 1in i=5,6
\eea
Here, $\g_m$ and $\g_i$ are the Dirac matrices on $AdS_5$ and $S^2$
respectively (see Appendix \ref{appA1}).
Integrability  requires $\eta _1^2 = \eta _2 ^2 =1$, and the
respective solution spaces are indeed of dimension 4 and 2, so that
$\chi ^{\eta _1, \eta _2}$ has 8 independent solutions.

\smallskip

The chirality matrix $\gamma_{(2)}$ of the $\g_i$-Dirac algebra produces
a reversal of the sign of $\eta_2$, allowing us to make the following identification,
\bea
\label{chiralities}
\gamma _{(2)} \chi ^{\eta _1, \eta _2} = \chi ^{\eta _1, - \eta _2}
\eea
Charge conjugation, under which $\chi \to \chi^c$, with
\bea
\left ( \chi ^c \right ) ^{\eta _1, \eta _2}  = (B_{(1)} \otimes B_{(2)})^{-1}
\left ( \chi ^{\eta _1 , \eta _2} \right )^*
\eea
reverses the signs of both $\eta _1$ and $\eta _2$, so that
$\left ( \chi ^c \right ) ^{\eta _1, \eta _2}$ is proportional to
$ \chi ^{-\eta _1 , -\eta _2}$. The proportionality factor depends on
$\eta_1, \eta_2$, and may be determined once a sign convention
has been chosen for one of the four components of $\chi ^{\eta _1, \eta _2}$. It
will be convenient to make the choice $\chi^{c++}= \chi^{--}$.
The remaining components are then related by using the $\g_{(2)}$-chirality
matrix, and the properties of the charge conjugation matrices $B_{(1)}$ and
$B_{(2)}$. For example, we have $\chi^{c+-} = (B_{(1)} \otimes B_{(2)})^{-1}
\g_{(2)} \chi^{*++} $ by the definition of $\chi^c$, and the action of  $
\gamma _{(2)}$. Commuting the $\g$-matrix past the $B$-matrices,
we have $\chi ^{c+-} = -\g_{(2)} (B_{(1)} \otimes B_{(2)})^{-1} \chi^{*++} $,
which in turn equals $ - \chi^{-+}$. The remaining relations may be similarly
obtained,  and we find,
\be
(\chi^c)^{\eta _1, \eta _2} = \eta_2 \chi ^{-\eta _1, -\eta _2}
\ee
a relation which is now valid for all four assignments of $\eta _1, \eta _2$.

\sm

The Killing spinors on $S^1$ are solutions to the equation,
\bea
(\hat \nabla _9 - {i \over 2} \eta_3 ) \chi^{\eta_3} = 0
\eea
Here,  we have $\eta_3 = \pm$, and $\chi^{\eta_3}$ consists of just a single
function, since the irreducible spinor representation of $SO(2)$ on $S^1$ are
1-dimensional. By choosing a proper overall phase convention, we may set
$(\chi^{\eta_3})^* = \chi^{-\eta_3}$. Note that the values of $\eta_3$ are
not required by integrability, since the above equation is integrable for all
values of  $\eta_3$. Rather, the values $\eta_3 = \pm 1$ naturally correspond to
a double-valued representation, suitable for spinors. Other odd values of
$\eta _3$ are allowed as well, but will be equivalent to our choice, upon redefinition
of the metric function $f_1$ of the circle $S^1$.

\sm

Putting all together, the full 32-component spinor $\ep$ may thus be decomposed
as follows in terms of Killing spinors on $AdS_5 \times S^2 \times S^1$,
\bea
\label{ep1}
\ep = \sum _{\eta _1, \eta _2,\eta_3}
\chi ^{\eta _1, \eta _2} \chi^{\eta_3}
\otimes
\l( \zeta _{\eta _1, \eta _2,\eta_3}
\otimes u_+  +  \zeta'_{\eta _1, \eta _2,\eta_3} \otimes u_-  \r)
\eea
where we have used the following basis,
\bea
\label{uu}
u_+ \equiv  { 1 \choose 0}  \hskip 1in u_- \equiv { 0 \choose 1}
\eea
Here, $\zeta _{\eta _1, \eta _2,\eta_3}$, and $\zeta'_{\eta _1, \eta _2, \eta_3}$
are independent 2-dimensional spinor functions of $\Sigma$.
The counting of components is as follows: $\chi^{\eta _1 \eta _2} $ has 8
components, $\chi^{\eta _3}$ has just one component, and the factor
in parentheses in (\ref{ep1}) a general 4-component spinor. The 10-dimensional
chirality condition $\G^{11} \ep = - \ep$ relates $\zeta'$ to $\zeta$ since we
have $\G^{11} = I_4 \otimes I_2 \otimes I_2 \otimes \sigma _2$, and thus,
\bea
\G^{11} \ep = \sum _{\eta _1, \eta _2,\eta_3}
\chi ^{\eta _1, \eta _2} \chi^{\eta_3}
\otimes
\l( i \zeta _{\eta _1, \eta _2,\eta_3}
\otimes u_-  -i   \zeta'_{\eta _1, \eta _2,\eta_3} \otimes u_+  \r)
\eea
As a result, we have $\zeta'_{\eta _1, \eta _2,\eta_3} = - i \zeta_{\eta _1, \eta _2,\eta_3}$.
By absorbing a constant phase into the $\zeta$ spinor, the last entry of the
tensor product in the 10-dimensional spinor
may now be recast in terms of a constant spinor $\theta$, defined by
\bea
\theta = e^{ i \pi /4} u_+ + e^{-i \pi /4} u_-
\eea
Its overall phase was chosen so that $\theta$ obeys $\sigma^1 \theta = \theta^*$,
$\s^2 \theta = -\theta$,  and $\sigma^3 \theta = i \theta^*$. The full 10-dimensional
spinor then takes the final form we shall use,
\bea
\ep = \sum _{\eta _1, \eta _2,\eta_3} \chi ^{\eta _1, \eta _2} \chi^{\eta_3} \otimes
 \zeta _{\eta _1, \eta _2,\eta_3} \otimes \theta
\eea
It will be convenient to use matrix notation
for the action of the indices $\eta_1, \eta _2, \eta _3$,
\bea
\label{tau3}
\tau ^{(ijk)}  \equiv  \tau^i \otimes \tau ^j \otimes \tau ^k \hskip 1in i,j,k =0,1,2,3
\eea
where $\tau^0=I_2$, and $\tau^i$ with $i=1,2,3$ are the  Pauli matrices in the standard
basis. Multiplication by $\tau ^{(ijk)}$ is defined as follows,
\bea
\label{tau}
(\tau ^{(ijk)} \zeta  )_{\eta _1 , \eta _2, \eta_3 }
 \equiv  \sum _{\eta _1 ', \eta _2 ',\eta_3'}
(\tau ^i)_{\eta _1 \eta _{1'}} (\tau ^j)_{\eta_{2} \eta_{2'}}
(\tau ^k)_{\eta_{3} \eta_{3'}}
\zeta _{\eta _1', \eta _2 ', \eta_3'}
\eea
Henceforth, we shall use matrix notation for $\tau$ and suppress the indices $\eta$.

\subsection{The reduced BPS equations}

Details of the reduction of the BPS-equations of (\ref{BPS}) have been deferred to 
Appendix \ref{appBPS}. The final results are as follows. The dilatino BPS equation is given by,
\bea
\label{dilatino1}
(d) \hskip 0.4in
0 =  4 p_a  \sigma^a \sigma^2 \zeta^*
+  g_{\tilde a} \tau^{(131)}  \sigma^{\tilde a} \zeta
+  h  \tau^{(121)}  \zeta
\eea
while the gravitino equation decomposes into a system of 4 equations,
\bea
\label{gravitino1}
(m) &\qquad&
0 = {i \over 2 f_5} \tau^{(300)} \zeta
+  {D_{a} f_5 \over 2 f_5} \tau^{(010)} \s^{a} \zeta
-{i \over 2}  f  \zeta
 \no\\&& \hskip 0.4in
+ {1 \over 16} \left ( i g_{\tilde a} \tau ^{(121)}\sigma^{\tilde a} \sigma ^2 \zeta^*
+  i h \tau^{(131)} \sigma ^2  \zeta^* \right )
\no\\
(i) &\qquad&
0 = {i \over 2 f_2}  \tau^{(030)}\zeta
+  {D_{a} f_2 \over 2 f_2} \tau^{(010)} \s^{a} \zeta
+ {i \over 2}  f  \zeta
\no\\&& \hskip 0.4in
+ {1 \over 16} \left ( -3 i g_{\tilde a}  \tau ^{(121)} \sigma^{\tilde a} \sigma ^2  \zeta^*
+ i h \tau^{(131)} \sigma ^2 \zeta^* \right )
\no\\
(9) &\qquad&
0 = {i \over 2 f_1}  \tau^{(013)} \s^3 \zeta
+  {D_{a} f_1 \over 2 f_1} \tau^{(010)} \s^{a} \zeta
 + {i \over 2}  f  \zeta
\no\\&& \hskip 0.4in
+ {1 \over 16} \left ( -3i g_9  \tau^{(121)}\sigma^{3}\s^2 \zeta^*
+ i g_a \tau^{(121)} \s^a \s^2\zeta^*
- 3i h \tau^{(131)}\s^2\zeta^* \right )
\no\\
(a) &\qquad&
0 =  (D_{a} + {i\over 2} \hat \omega_{a} \s^3) \zeta
- {i q_{a} \over 2}    \zeta + {i f\over 2}\tau^{(010)} \s_a   \zeta
\no\\&& \hskip 0.4in
+ {1 \over 16} \left ( 3 g_a \tau^{(131)} \sigma ^2  \zeta^*
- g_{\tilde b} \tau^{(131)}\s_a {}^{ \tilde b} \sigma ^2 \zeta^*
- 3 h  \tau ^{(121)} \sigma_a \sigma ^2 \zeta^* \right )
\eea

\subsection{Symmetries of the reduced BPS equations}

The symmetries of the reduced BPS equations will play a fundamental role
in the analysis of half-BPS solutions. It will be natural to separate
the symmetries into three groups: continuous symmetries;
{\sl linear } discrete symmetries which map $\zeta$ to a linear function
of $\zeta$; and complex conjugations which map $\zeta $ to a linear
function of $\zeta ^*$.

\subsubsection{Continuous symmetries}

One manifest continuous symmetry consists of the $U(1)$ frame rotations
on the orthogonal frame $e^a$ on $\Sigma$.

\sm

The axion/dilaton field $B$ transforms non-linearly under the continuous
$S$-duality group  $SU(1,1)$ of Type IIB supergravity, and the field $B$
takes values in the coset $SU(1,1)/U(1)_q$. Thus, $SU(1,1)$ transformations
on the $B$-field and on other fields are accompanied by local $U(1)_q$ gauge transformations, given in (\ref{su11a}) and (\ref{su11b}). This $U(1)_q$
survives the reduction to our Ansatz, and induces the following
transformations,
\bea
\label{U1qsymmetry}
\zeta \to e^{i \theta/2} \zeta \hskip 0.27in  & \hskip .6in & g_a \to e^{ i \theta} g_a
\no \\
q_a \to q_a +  D_a \theta    & \hskip .6in & g_9 \to e^{ i \theta} g_9
\no \\
p_a \to e^{2 i \theta} p_a  \hskip 0.25in  & \hskip .6in &  h \, \to e^{ i \theta} h
\eea
The real function $\theta$ depends on the  $SU(1,1)$ transformation,
as well as on the field $B$.

\subsubsection{Linear discrete symmetries}

The following linear transformations on $\zeta$ are symmetries of all
the BPS equations, upon leaving the metric factors $f_1, f_2, f_5$,
the flux fields $f, g_a, g_9, h, p_a, q_a$ unchanged,
\bea
\zeta \to \zeta ' = S \zeta
\hskip 1in
S \in \cS_0 \equiv \left \{ I, \tau^{(303)}, i \tau^{(300)}, i \tau ^{(003)} \right \}
\eea
The set $\cS_0$ forms an Abelian group.
The BPS equations have further linear symmetries $\zeta \to \zeta ' = S \zeta$,
given by,
\bea
S \in \cS_1 \equiv \left \{
\tau^{(331)} \sigma^3, \,
\tau^{(332)} \sigma^3, \,
i \tau^{(031)} \sigma^3, \,
i \tau^{(032)} \sigma^3 \right \}
\eea
The transformations in $\cS_1$ leave $f_1, f_2, f_5$, and
$f, g_a, h, p_a, q_a$ unchanged but, in contrast with the
transformations in $\cS_0$, must be accompanied by a sign reversal
$g_9 \to g_9'=-g_9$. Thus, the generators of $\cS_1$ are genuine
symmetries of the bosonic supergravity  fields only if $g_9=0$.
The  union $\cS_0 \cup \cS_1$
forms a non-Abelian group, whose generators may be chosen to be
\bea
\label{st1}
T_1 \zeta  & = & \tau ^{(303)} \zeta
\no \\
T_2 \zeta  & = & i \tau ^{(300)} \zeta
\no \\
T_3 \zeta & = & \tau ^{(331)} \sigma ^3\zeta
\eea

\subsubsection{Complex conjugation (or anti-linear) symmetries}

The BPS equations are invariant under 8 discrete complex conjugations,
which may be viewed as the compositions of the 8 transformations of
$\cS_0 \cup \cS_1$ with a single complex
conjugation. It will be useful to record a more general transformation,
in which we compose the discrete complex conjugation with an arbitrary
continuous $U(1)_q$ transformation, which gives,
\bea
\label{cc}
\zeta \to T_4 \zeta = \zeta ' = e^{ i \theta} \tau ^{(030)} \sigma ^1 \zeta ^*
& \hskip 0.7in &
g_a \to g_a '=  e^{2i\theta} \, g^*_a
\no \\ &&
g_9 \to g_9' = - e^{2i \theta } \, g^*_9
\no \\ &&
h \, \to h' =  e^{2i\theta} \,  h^*
\no \\ &&
p_a \to p'_a =  e^{4 i\theta} \,  p^*_a
\no \\ &&
q_a \to q_a'=  -q_a + 2 D_a\theta
\eea
The pure complex conjugation corresponds to the special case where $\theta =0$.

\subsection{Further reduction using $\zeta \to \tau ^{(303)} \zeta$ symmetry }

The transformations in $\cS_1$ are genuine symmetries of the BPS
equations only if $g_9=0$, while the complex conjugations are
symmetries only if the dilaton/axion and flux fields obey reality
conditions $ p_a^* = p_a$, $q_a=0$, $ g_a^*=g_a$, $ g_9^* = - g_9$,
and $h^* = h$, up to $U(1)_q$. For general bosonic fields, only
the transformations in $\cS_0$ are genuine symmetries of the BPS equations.
Only the transformation $\zeta \to \tau ^{(303)} \zeta$ commutes
with the BPS differential operator acting on real linear combinations,
so that this is the only symmetry that may be diagonalized simultaneously
with the BPS operator. (Note that the other two non-trivial symmetries
of $\cS_0$ square to $-I$, and admit no real eigenvalues.) Thus, we
may now analyze separately the restrictions of the BPS equations
to the two eigenspaces of $\tau^{(303)}$, given $\nu = \pm 1$,
\bea
\tau ^{(303)} \zeta = \nu \, \zeta
\eea
Redefining the non-vanishing component of the spinor  $\zeta$
in terms of a spinor with only two indices (which, by abuse of notation,
we denote again by $\zeta$), we have,
\bea
\nu = 1
\left\{
\begin{array}{c}
\zeta_{++} \equiv \zeta_{+++}
\no \\
\zeta_{+-} \equiv \zeta_{+-+}
\no \\
\zeta_{-+} \equiv \zeta_{-+-}
\no \\
\zeta_{--} \equiv \zeta_{---}
\no \\
\end{array}
\right.
& \hskip 0.8in &
\nu = -1
\left\{
\begin{array}{c}
\zeta_{++} \equiv \zeta_{++-}
\no \\
\zeta_{+-} \equiv \zeta_{+--}
\no \\
\zeta_{-+} \equiv \zeta_{-++}
\no \\
\zeta_{--} \equiv \zeta_{--+}
\no \\
\end{array}
\right.
\eea
In terms of this two-index spinor $\zeta$, the BPS equations become,
\bea
\label{BPS2}
(d) &\qquad&
0 =  4 p_a  \sigma^a \sigma^2 \zeta^*
+  g_{\tilde a} \tau^{(13)}  \sigma^{\tilde a } \zeta
+  h  \tau^{(12)}  \zeta
\no \\ \no \\
(m) &\qquad&
0 = {i \over 2 f_5} \tau^{(30)} \zeta
+  {D_{a} f_5 \over 2 f_5} \tau^{(01)} \s^{a} \zeta
-{i \over 2}  f  \zeta
\no\\&& \hskip 0.4in
 + {1 \over 16} \Big ( i g_{\tilde a} \tau ^{(12)}\sigma^{\tilde a } \sigma ^2 \zeta^*
 +  i h \tau^{(13)} \sigma ^2  \zeta^* \Big )
\no\\ \no \\
(i) &\qquad&
0 = {i \over 2 f_2}  \tau^{(03)}\zeta + {D_{a} f_2 \over 2 f_2} \tau^{(01)} \s^{a} \zeta
+ {i \over 2}  f  \zeta
\no\\&& \hskip 0.4in
+ {1 \over 16} \Big ( -3 ig_{\tilde a }  \tau ^{(12)} \sigma^{\tilde a} \sigma ^2  \zeta^*
+ i h \tau^{(13)} \sigma ^2 \zeta^* \Big )
\no\\ \no \\
(9) &\qquad&
0 = {i \nu\over 2 f_1}  \tau^{(31)}\s^3\zeta +  {D_{a} f_1 \over 2 f_1} \tau^{(01)} \s^{a} \zeta
 + {i \over 2}  f  \zeta
\no\\&& \hskip 0.4in
+ {1 \over 16} \Big (
i g_a \tau^{(12)} \s^a \s^2\zeta^*
- 3 i g_9 \tau ^{(12)} \s^3 \s^2 \zeta ^*
- 3i h \tau^{(13)}\s^2\zeta^* \Big )
\no\\ \no \\
(a) &\qquad&
0 =  (D_{a} + {i\over 2} \hat \omega_{a} \s^3) \zeta
- {i q_{a} \over 2}    \zeta + {i f\over 2}\tau^{(01)} \s_a   \zeta
\no\\&& \hskip 0.4in
+ {1 \over 16} \Big ( 3 g_{a} \tau^{(13)} \sigma ^2  \zeta^*
- g_{\tilde b} \tau^{(13)}\s_a {}^{ \tilde b} \sigma ^2 \zeta^*
- 3 h  \tau ^{(12)} \sigma_a \sigma ^2 \zeta^* \Big )
\eea
The matrices $\tau ^{(ij)}$ act on the two-index spinor $\zeta$ in the
obvious manner.

\sm

Remarkably, the dependence on $\nu$ of the reduced BPS equations
is entirely concentrated in the first term of equation (9) in (\ref{BPS2}),
and may be undone by changing the sign of the metric factor $f_1$.
As only the square $f_1^2$ enters into any of the supergravity fields,
this sign redefinition is immaterial, and the projections $\nu =\pm 1$
represent the same bosonic supergravity fields. A solution
to the reduced BPS equations for $\nu =+1$ thus automatically
produces another  solution with the same bosonic fields but with $\nu=-1$.
In this manner, a systematic doubling of the number of spinor
solutions is produced.

\sm

As a corollary, this doubling up of the number of spinor solutions implies
that any solution to the BPS equations with $\nu=+1$ effectively produces 16
linearly independent solutions of the full BPS equations, and thus generates
a half-BPS solution.

\subsection{Residual symmetries of the reduced form}

Having  projected $\zeta$ onto the two eigenspaces $\tau ^{(303)} \zeta = \nu \zeta$,
any residual symmetry of  (\ref{BPS2}) must map these eigenspace into themselves,
requiring the transformation to commute with $\tau^{(303)}$.
The transformations $U(1)_q$  thus continue to be symmetries of  (\ref{BPS2}).
The remaining elements in $\cS_0$, namely $i \tau ^{(300)}$
and $i \tau ^{(003)}$ map between identical $\nu$, and reduce to the following
transformations on two-index $\zeta$ variables,
\bea
i \tau ^{(300)} & \qquad \to \qquad & i \tau ^{(30)}
\no \\
i \tau ^{(003)} & \qquad \to \qquad & i \nu \tau ^{(30)}
\eea
both of which are symmetries of the further reduced BPS equations (\ref{BPS2}).
The transformations in $\cS_1$ all anti-commute with $\tau ^{(303)}$,
thus map the system with $\nu$ to the system with $-\nu$, and do not
yield symmetries of the further reduced BPS equations (\ref{BPS2}).

\sm

Amongst the complex conjugation symmetries of the original reduced
BPS equations (\ref{dilatino1}) and (\ref{gravitino1}), only the transformation
of (\ref{cc}), and its images under $\cS_0$, commute with $\tau^{(303)}$.
The transformation of (\ref{cc}) acting on 2-index $\zeta$ variables is given by,
\bea
\label{cc2}
\zeta \to \zeta ' = e^{i \theta } \tau ^{(03)} \sigma ^1 \zeta^*
\eea
and is a residual symmetry of the further reduced BPS equations (\ref{BPS2}).

\subsection{Chiral form of the reduced BPS equations}

It will be convenient to separate the two components of the
2-dimensional spinors $\zeta _{\eta _1, \eta _2, \eta _3}$, or of
their two-index reductions $\zeta _{\eta _1, \eta _2}$. This
decomposition goes hand in hand with the complex structure
decomposition on the Riemann surface $\Sigma$, and may be achieved
by introducing the standard complex frame basis $e^a = (e^z, e^{\bar z})$,
with metric $\delta _{z \bar z} = \delta _{\bar z z} = 2$. The explicit
expressions for the frame and flux fields in this basis may be deduced
from
\bea
\label{complex}
e^z = (e^7 + i e^8)/2
& \hskip 1in &
e_z = e^7 - i e^8
\no \\
e^{\bar z} = (e^7 - i e^8)/2
& \hskip 1in &
e_{\bar z} = e^7 + i e^8
\eea
Similar relations hold for the components of the fields $p_a, q_a, g_a$,
so that, for example, $p_z = p_7 - i p_8$, and $p_{\bar z} = p_7 + i p_8$.
The Pauli matrices in this basis take the form,
\bea
\sigma ^z = \left ( \matrix{ 0 & 1 \cr 0 & 0 \cr} \right )
\hskip 1in
\sigma ^{\bar z} = \left ( \matrix{ 0 & 0 \cr 1 & 0 \cr} \right )
\eea
We proceed to decompose the two-index spinor $\zeta$ into its two
chirality components in this same 2-dimensional spinor basis,
\be
\zeta_{\eta_1,\eta_2} = {\tau^{(13)} \, \xi^*_{\eta_1,\eta_2} \choose \psi_{\eta_1,\eta_2}}
\ee
where $\xi ^*_{\eta_1,\eta_2}$ and $\psi _{\eta_1,\eta_2}$ are
1-component spinors. In this basis, the reduced BPS equations take the following form,
\bea
\label{BPS2a}
(d_1) &\qquad&
4 i p_z  \xi
+  g_{z}  \psi
+ g_9 \tau ^{(13)} \xi ^*
-  h  \tau^{(12)}  \xi^* = 0
\no \\
(d_2) &\qquad&
 4 i p^*_{\bar z}  \psi
+  g^*_{\bar z}  \xi
- g^*_9 \tau ^{(13)} \psi^*
-  h^*  \tau^{(12)}  \psi^* = 0
\no \\
\no \\
(m_1) &\qquad& {i \over  f_5} \tau^{(22)} \xi^* + {D_{z} f_5 \over
f_5}   \psi -   f \tau^{(12)} \xi^*
 + {i \over 8} ( -  g_{z} \xi + g_9 \tau ^{(13)} \psi ^* - h \tau^{(12)} \psi^*) = 0
\no\\
(m_2) &\qquad&
-{i \over  f_5} \tau^{(22)} \psi^*
+  {D_{ z} f_5 \over  f_5}   \xi
- f \tau^{(12)}    \psi^*
 + {i \over 8} (   -g^*_{\bar z} \psi - g^*_9 \tau ^{(13)} \xi^* -  h^* \tau^{(12)} \xi^*) = 0
\no\\ \no \\
(i_1) &\qquad&
 {i \over  f_2}  \tau^{(11)}\xi^* +  {D_{z} f_2 \over  f_2}  \psi
+    f \tau^{(12)}  \xi^*
+ {i \over 8} ( 3 g_{z} \xi -3 g_9 \tau ^{(13)} \psi ^* -  h \tau^{(12)} \psi^*) =0
\no\\
(i_2) &\qquad&
{i \over  f_2}  \tau^{(11)}\psi^* +  {D_{z} f_2 \over  f_2}  \xi
+    f \tau^{(12)} \psi^*
+ {i \over 8} ( 3 g^*_{\bar z} \psi + 3 g^*_9 \tau ^{(13)} \xi^* - h^* \tau^{(12)} \xi^*) =0
\no\\ \no \\
(9_1) &\qquad&
 -{\nu\over  f_1}  \tau^{(23)}\xi^* +  {D_{z} f_1 \over  f_1}  \psi
 + f \tau^{(12)} \xi^*
+ {i \over 8} ( - g_z \xi -3 g_9 \tau ^{(13)}\psi ^* + 3 h \tau^{(12)}\psi^*)=0
\no \\
(9_2) &\qquad&
 { \nu \over  f_1}  \tau^{(23)}\psi^* +  {D_{ z} f_1 \over  f_1}  \xi
+ f \tau^{(12)} \psi^*
+ {i \over 8} ( - g^*_{\bar z} \psi + 3 g^*_9 \tau ^{(13)} \xi^* + 3  h^* \tau^{(12)}\xi^*)=0
\no \\ \no \\
(+_1) &\qquad&
\left  ( D_{\bar z} - {i\over 2} \hat \omega_{\bar z} + {i  \over 2} q_{\bar z} \right )  \xi
+ {i \over 4}  g^*_{z} \psi = 0
\no \\
(+_2) &\qquad&
\left ( D_{z} - {i\over 2} \hat \omega_{z} - {i  \over 2} q_{z} \right )  \psi
+ {f } \t^{(12)} \xi^*
+ {i \over 8} (  g_{z}  \xi  + g_9 \tau ^{(13)} \psi ^* + 3 h\t^{(12)} \psi^*)= 0
\no\\ \no \\
(-_1) &\qquad&
\left ( D_z - {i\over 2} \hat \omega_z + {i  \over 2} q_z \right )  \xi
+ f  \t^{(12)} \psi^*
+ {i \over 8} (  g_{\bar z}^*  \psi - g_9^* \tau ^{(13)} \xi^*  + 3 h^* \t^{(12)} \xi ^* ) = 0
\no\\
(-_2) &\qquad&
\left  ( D_{\bar z} - {i\over 2} \hat \omega_{\bar z}  - {i  \over 2} q_{\bar z} \right )   \psi
+ {i \over 4}  g_{\bar z}  \xi = 0
\eea
where $\hat \o_z = i (\p_w \rho )/\rho^2$.
The action of the complex conjugation symmetry
(\ref{cc2}) is given by
\bea
\label{cc3}
\xi & \to & \xi ' = e^{- i \theta} \tau ^{(10)} \psi
\no \\
\psi & \to & \psi ' = e^{+ i \theta} \tau ^{(10)} \xi
\eea
It acts by interchanging $\xi$ and $\psi$, together with the
transformations on the bosonic fields, listed already in (\ref{cc}),
and translated as follows in the chiral basis,
\bea
\label{cc4}
p_z \to p_z' = e^{4 i \theta } (p_{\bar z} )^* \hskip 0.27in
& \hskip 0.6in &
g_z  \to  g_z' = e^{2 i \theta } (g_{\bar z} )^* \hskip 0.6in 
h \to h' = e^{2 i \theta }  h ^*
\no \\
q_z  \to  q_z' = - q _z + 2 D_z \theta
& \hskip 0.6in &
g_9  \to  g_9' = - e^{2 i \theta } g_9^*
\eea
This concludes the full reduction of the BPS equations onto the
$AdS_5 \times S^2 \times S^1 \times \Sigma$ Ansatz.

\section{Vanishing G implies the $AdS_5 \times S^5$ solution}
\setcounter{equation}{0}
\label{four}

In this section, we shall prove that, whenever the 3-form flux field $G$ vanishes,
 the solution to the BPS equations
is $AdS_5 \times S^5$ itself.  We shall also spell out the properties of this solution.

\sm

When $G=0$, we have $g_z=g_{\bar z} = g_9 = h = 0$. For solutions with
at least 16 supersymmetries,   $\psi$ and $\xi$ cannot both vanish.
The reduced dilatino equations (d) in (\ref{BPS2a}) imply 
$p_z = p_{\bar z}=0$, so that $P=0$. By the Bianchi identities, we then have $dQ=0$.
Using the $U(1)_q$ gauge symmetry (\ref{U1qsymmetry}), we may now choose a
gauge in which $Q=0$, so that,
\bea
\label{S1}
 g_z = g_{\bar z} = g_9 = h = 0
\no \\
p_z = p_{\bar z} = q_z = q_{\bar z} = 0
\eea
The gauge choice leaves open the symmetry of performing constant gauge
transformations, which we shall fix later.

\subsection{Using the discrete symmetries of the reduced BPS equations}

Given the vanishing of the components of $G,P,Q$ in (\ref{S1}), all four
generators $T_1, T_2, T_3$, and $T_4$ of (\ref{st1}) and (\ref{cc})
produce genuine discrete symmetries  of (\ref{BPS2a}). The generators
$T_1, T_2$, and $T_4$ mutually commute and may be simultaneously
diagonalized as follows,
\bea
T_1 \zeta & = & \t^{(303)} \zeta  = \nu \zeta
\no \\
T_2 \zeta & = & \t^{(300)} \zeta = \gamma \zeta
\no \\
T_4 \zeta & = & e^{ i \theta} \tau ^{(030)} \sigma ^1 \zeta ^* = \mu \zeta
\eea
where $ \nu, \gamma, \mu$ take on independently the values $\pm1$.
The projection of $T_1$ has already
been carried out in the previous section. Clearly, the sign $\mu$ in the $T_3$ projection
can always be gauged away with a constant gauge transformation in $U(1)_q$.
Henceforth, we shall choose $\mu e^{i \theta }=1$. In terms of
$\psi$ and $\xi$, these projections become,
\bea
T_2: &\qquad &
\tau ^{(30)} \psi   =  \g \psi
 \qquad
\tau ^{(30)} \xi   =  -\g \xi
\no \\
T_4: &\qquad &
\xi = \t^{(10)} \psi
\eea
Solutions with opposite values of $\g$ are equivalent.  Henceforth, we set $\gamma =+$,
and shall treat $\psi$ and $\xi$ as two-component spinors, omitting their fixed $\eta_1$
indices.
\bea
\psi & = &{ \psi_{+} \choose \psi_{-}} \equiv { \psi_{++} \choose \psi_{+-}}
\no \\
\xi & =& { \xi_{+} \choose \xi_{-}} \equiv { \xi_{-+} \choose \xi_{--}} = { \psi_{++} \choose \psi_{+-}}
\eea
After these simplifications, the remaining reduced BPS
equations of (\ref{BPS2a}) take the form,
\bea
\label{BPS5}
(m) &\qquad&
 \mp{i \over  f_5} \psi^*_\mp  + {D_{z} f_5 \over  f_5}   \psi_\pm  \pm  i f  \psi^*_\mp  = 0
\no\\
(i) &\qquad&
 {i \over  f_2}  \psi^*_\mp  + {D_{z} f_2 \over  f_2}   \psi_\pm  \mp i f  \psi^*_\mp  = 0
\no\\
(9) &\qquad&
\pm{i \nu \over  f_1}  \psi^*_\pm  + {D_{z} f_1 \over  f_1}   \psi_\pm  \mp i f \psi^*_\mp  = 0
\no \\
(-) &\qquad&
(D_{\bar z} - {i\over 2} \hat \omega_{\bar z} ) \psi_\pm = 0
\no \\
(+) &\qquad&
 (D_{z} - {i\over 2} \hat \omega_{z} ) \psi_\pm  \mp i  f  \psi^*_\mp = 0
\eea

\subsection{Generic solutions when $G=0$}

The $(m)$-equations may be expressed  as follows,
\bea
\label{mm2}
{D_{z} f_5 \over  f_5}  \psi_\pm & = & \pm i \left ( { 1\over  f_5} -  f  \right ) \psi^*_\mp
\eea
The product of the $+$ equation with the complex conjugate of the $-$ equation gives,
\bea
\label{mm3}
 \left|{D_{z} f_5 \over  f_5} \right|^2 \psi_+ \psi^*_- + \left ( { 1\over  f_5} -  f \right )^2 \psi_+ \psi^*_- =0
\eea
When $\psi _+ \psi _-^*\not=0$,  (\ref{mm3}) implies $\p_z f_5=0$ and $ff_5=1$.
If $\psi _+ \psi _- ^*=0$, then either $\psi _+=0$ or $\psi _-^*=0$, but both cannot vanish
simultaneously.
In either case, (\ref{mm2}) implies $\p_z f_5=0$ and $ff_5=1$. Thus, $f$ and $f_5$
are constant satisfying $ff_5=1$.

\sm

$\bullet$ Next, using the expression $\hat \o_z = i (\p_z \rho )/\rho^2$, and
$D_z = \rho ^{-1} \p_z $, we  derive the complete solution to the $(-)$ equations
of (\ref{BPS5}), as follows,
\bea
\label{sol-}
\psi _+ = \sqrt{\rho} \,\alpha , \qquad \psi_- = \sqrt{\rho}
\,\beta & \hskip 1in & \p_{\bar z} \a = \p_{\bar z} \b =0
\eea
so that $\a$ and $\b$ are holomorphic functions.

\sm

$\bullet$ From the differential equations $(\pm)$ in (\ref{BPS5}), we find
$D_z(|\psi _+|^2 + |\psi _-|^2) = 0$. Therefore $|\psi _+|^2 + |\psi _-|^2$
is a constant; we may normalize this to be $f_5$, which is also a constant.
\bea
\label{f5}
f _5 =   |\psi _+|^2 + |\psi _-|^2 = \rho (\a \bar \a + \b \bar \b)
\eea

\sm

$\bullet$
Using equation (\ref{f5}) to eliminate $f=f_5^{-1}$ from the $(+)$ equation in (\ref{BPS5}),
we find,
\bea
\label{BPS-AdS5+}
\p_z \left ( { \a \over \a \bar \a + \b \bar \b} \right )
- {  i  \, \bar \b \over (\a \bar \a + \b \bar \b)^2} & = & 0
\no \\
\p_z \left ( { \b \over \a \bar \a + \b \bar \b} \right )
+  { i  \, \bar \a \over (\a \bar \a + \b \bar \b)^2} & = & 0
\eea
Multiplying the first by $\bar \a$, the second by $\bar \b$, and adding gives
an equation that is automatically satisfied. Thus, we may retain just the first
equation. After some simplifications, it reduces to a holomorphic equation,
\bea
\b \p_z \a - \a \p_z \b - i =0
\eea
which may be easily solved, and gives the solution
\bea
\beta (z) =  -i  \a (z) A(z) \hskip 1in A(z) \equiv \int ^z_{z_0} { du \over \a (u)^2}
\eea
with $z_0$ being the arbitrary integration constant. In fact, it will be
convenient to express both $\a$ and $\b$ solely in terms of the function $A$,
namely,
\bea
\a (z) = { 1 \over \sqrt{\p_z A(z)}} \hskip 1in \beta (z) = -{ i  A(z) \over \sqrt{\p_z A(z)}}
\eea
To determine metric factors in terms of spinor components\footnote{A complete
analysis of Hermitian forms will be given in the next section.} we make use of both
algebraic and differential equations. On the one hand, from the (i) equations in
(\ref{BPS5}),  we have
\bea
{D_z f_2 \over f_2} ( |\psi_+|^2 -|\psi_-|^2 ) = 2if \psi_-^* \psi_+^*
\eea
On the other hand, using differential equations ($+$) and ($-$), we  obtain the relation
\bea
D_z ( |\psi_+|^2 -|\psi_-|^2 ) = 2if \psi_-^* \psi_+^*
\eea
Combining the above equations gives,
\bea
f_2 = c_2 (|\psi_+|^2 -|\psi_-|^2 )
\eea
where $c_2$ is a constant. Similarly, from (9) equations, we may derive the following relations,
\bea
{D_z f_1 \over f_1} ( \psi_+^* \psi_-  + \psi_-^* \psi_+ ) = if ( \psi_-^* \psi_-  - \psi_+^* \psi_+ )
\no \\
D_z ( \psi_+^* \psi_-  + \psi_-^* \psi_+ ) = if ( \psi_-^* \psi_-  - \psi_+^* \psi_+ )
\eea
Combining these in turn implies,
\bea
 f_1 = c_1 ( \psi_+^* \psi_-  + \psi_-^* \psi_+ )
\eea
where $c_1$ is a constant.

$\bullet$ Using the expression for metric factors, $\a$ and $\beta$, it can be checked that
(i) and (9) equations are all satisfied, if $c_1 = c_2 =1 $. So all the reduced BPS equations
are completely solved, and all fields are determined by a holomorphic function $A(z)$.

\subsection{Solution of $AdS_5 \times S^5$}

Choosing $A(z)=- e^{-2z}, f_5 = 1$, we recover the $AdS_5 \times S^5$ solution,
\bea
d s^2  =  d s^2 _{AdS_5} + (\th \, x_7)^2 d s^2 _{S^2} +
{dx_7^2 + dx_8^2 \over (\ch \, x_7)^2} + { \sin^2x_8 dx_9^2 \over ( \ch \, x_7)^2}
\eea
To exhibit the metric in standard form, we change parametrization,
 $ e^{x^7} = \tan({\theta \over 2})$,
\bea
d s^2 & =&  d s^2 _{AdS_5} + \cos^2\theta d s^2 _{S^2} +
\sin^2\theta (dx_7^2 + dx_8^2) + \sin^2\theta \sin^2x_8 dx_9^2
\no \\
& =&  d s^2 _{AdS_5} + \cos^2\theta d s^2 _{S^2} +
d\theta^2 + \sin^2\theta dx_8^2 + \sin^2\theta \sin^2x_8 dx_9^2
\no \\
& =&  d s^2 _{AdS_5} + d\theta^2  + \cos^2\theta d s^2 _{S^2} +
\sin^2\theta d s^2 _{S^2}
\eea
which is indeed $AdS_5 \times S^5$.
The solution to the spinor $\zeta$ is characterized by three projections,
\bea
\tau^{(030)}\s^1 \zeta^* & = & \zeta
\no \\
\tau^{(303)} \zeta & = & \nu \zeta
\no \\
\tau ^{(300)} \zeta & = &  \gamma \zeta
\eea
where we can have, independently, $\nu = \pm 1, \gamma = \pm 1$.
In terms of functions $\zeta_1, \zeta_2$ defined below, the non-vanishing components
of spinor solutions are (with $z = (x_7 + i x_8)/2$):
\bea
\zeta_1 &= & i^{1- \nu \over 2} i^{1- \gamma \over 2} { e^{\bar z} \over \sqrt{2\ch \, x_7}}
\no \\
\zeta_2 &= &  i^{ \nu -1\over 2} i^{1+ \gamma \over 2}  { e^{- \bar z} \over \sqrt{2\ch \, x_7}}
\eea
\bea
\nu =+ 1, \quad \gamma = +1 &\hskip 0.2in& 
\zeta_{+++} = {\zeta_1\choose \bar\zeta_1} \qquad 
\zeta_{+-+} = {\zeta_2 \choose - \bar\zeta_2}
\no \\
\nu =+ 1, \quad \gamma = -1 && 
\zeta_{-+-} = {\zeta_1 \choose \bar\zeta_1} \qquad 
\zeta_{---} = {\zeta_2 \choose - \bar \zeta_2}
\no \\
\nu = -1, \quad \gamma = +1 && 
\zeta_{++-} = {\zeta_1 \choose \bar\zeta_1} \qquad 
\zeta_{+--} = {\zeta_2 \choose - \bar\zeta_2}
\no \\
\nu = -1, \quad \gamma = -1 && 
\zeta_{-++} = {\zeta_1 \choose \bar\zeta_1} \qquad 
\zeta_{--+} = {\zeta_2 \choose -\bar \zeta_2}
\eea
With 8 independent solutions to killing spinors $\chi$ and 4 independent solutions to
$\zeta$, we indeed recover 32
supersymmetries as expected.

\bigskip

\section{Metric factors in terms of spinor bilinears}
\setcounter{equation}{0}

The metric factors $f_1, f_2, f_5$ may be related to bilinear (or more
accurately Hermitian forms) of the spinors $\psi, \xi$. Since the
metric factors $f_1, f_2, f_5$ are real, the spinor bilinears need to be
real, and invariant under $U(1)_q$ transformations. These requirements
rule out combinations of the form $\psi ^t \tau ^{(\a \b)} \psi$, $\xi ^t \tau ^{(\a \b)} \xi$,
and $\psi ^t \tau ^{(\a \b)} \xi$, because they are complex, and
$\psi ^\dagger \tau ^{(\a \b)} \xi$, and  $\xi ^\dagger \tau ^{(\a \b)} \psi$
because they fail to be $U(1)_q$-invariant. Thus, the only real and
$U(1)_q$-invariant combinations left are $\psi ^\dagger \tau ^{(\a \b)} \psi$
and $ \xi ^\dagger \tau ^{(\a \b)} \xi$.

\subsection{Example of metric factor calculation}

The detailed calculation is relegated to appendix \ref{metric}, here, we shall illustrate the
procedure by giving an specific example of finding spinor bilinears of $f_5$.

\sm

We shall use combinations of the differential equations $(\pm)$ in (3.28)
and of the algebraic gravitino BPS equations to bring out the corresponding
relations. To this end, we compute,
\bea
D_z \left (\psi^{\dagger}  \psi \right )
&=&
-  f  \psi^{\dagger} \tau^{(12)}\xi^*
+ {i \over 4}  g_{\bar z}^* \xi^\dagger  \psi
- {i \over 8} \left (  g_{z} \psi^{\dagger}  \xi
+  g_{9}\psi^{\dagger} \tau^{(13)} \psi^*
+ 3h \psi^{\dagger} \t^{(12)}\psi^* \right )
\no \\
D_{z} \left ( \xi^{\dagger}   \xi \right )
& = &
+  f \psi^{\dagger} \tau^{(12)}   \xi^*
+ {i \over 4}  g_{z}  \psi^{\dagger}\xi
+ {i \over 8} \left ( - g_{\bar z}^*  \xi^{\dagger}   \psi
+  g^*_{9}\xi^{\dagger}  \tau^{(13)} \xi^*
- 3h^*\xi^{\dagger}  \t^{(12)} \xi^* \right )
\eea
Then, left-multiplying equation $(m_1)$  by $\psi^\dagger$, and left-multiplying
the complex conjugate of equation $(m_2)$ by $\xi ^\dagger$, we find,
\bea
{D_z f_5 \over f_5} \left ( \psi^{\dagger}   \psi \right )
& = &
- {i \over f_5} \psi^{\dagger}  \t^{(22)} \xi^*
+  f \psi^{\dagger} \t^{(12)} \xi^*
+{i \over 8} \left (  g_z \psi^{\dagger} \xi
-  g_{9} \psi^{\dagger}  \tau ^{(13)} \psi^*
+ h \psi^{\dagger}  \t^{(12)} \psi^* \right )
\no \\
{D_z f_5 \over f_5} \left ( \xi^{\dagger}  \xi \right )
& = &
+ {i \over f_5} \xi^{\dagger}  \t^{(22)} \psi^*
+ f \xi^{\dagger}\t^{(12)} \psi^*
+ {i \over 8} \left (  g^*_{\bar z} \xi^{\dagger}   \psi
+   g^*_{9} \xi^{\dagger}  \tau ^{(13)} \xi^*
+  h^* \xi^{\dagger}  \t^{(12)} \xi^* \right )
\eea
With appropriate combinations of the above equations, all terms would cancel out.
We may get the following relation,
\bea
\label{cons1}
D_z \left (\psi^{\dagger}  \psi +  \xi^{\dagger}   \xi \right )
- {D_z f_5 \over f_5} \left (  \psi^{\dagger} \psi +
 \xi^{\dagger}   \xi \right ) =0
\eea
Which further implies, $ D_z  [ f_5^{-1} ( \psi^{\dagger} \psi +
 \xi^{\dagger}   \xi) ] =0$, so that $f_5$ may be expressed as the product of
 constant $C_5^{(00)}$ and spinor bilinears,
\bea
f_5^{-1} \left( \psi^{\dagger} \psi +
 \xi^{\dagger}   \xi \right )
=   C_{5}^{(00)}
\eea

\subsection{Summary of metric factor expressions}

As a summary, we list all the {\sl generic relations}, valid for arbitrary
values of all the fields of the supergravity Ansatz.
For notational simplicity of this and the remaining sections, we shall use the following definition
for Hermitian forms.
\bea
\label{defH}
H_{\pm}^{(\a \b)} \equiv \psi^{\dagger} \t^{(\a \b)} \psi \pm \xi^{\dagger} \t^{(\a \b)} \xi
\eea
In terms of these Hermitian composites, we have the following result,
\bea
\label{gen1}
f_5^{-1} H_{+}^{(\a \b)}
=   C_{5}^{(\a \b)}
& \hskip 0.7in &
\t^{(\a \b)}  \in \{  \t^{(00)}, \t^{(31)} ,\t^{(32)}, \t^{(33)} \}
\no \\
f_5^3 H_{-}^{(\a \b)}
=   C_{5}^{(\a \b)}
&  &
\t^{(\a \b)} \in \{  \t^{(01)}, \t^{(30)} \}
\no \\ && \no \\
f_2^{-1} H_{-}^{(33)}
=   C_2^{(33)}
&&
\\
f_2^{1/3} H_{+}^{(\a \b)}
=   C_2^{(\a \b)}
&  &
\t^{(\a \b)}  \in \{ \t^{(31)},\t^{(32)} \}
\no \\ && \no \\
f_1^{-1} H_{+}^{(\a \b)}
= C_1^{(\a \b)}
& &
\t^{(\a \b)} \in \{  \t^{(01)},\t^{(30)},\t^{(32)},\t^{(33)} \}
\no
\eea
When $g_9=0$, we have additional non-generic relations, which are
summarized below,
\bea
\label{zerog}
f_5^3 H_{-}^{(\a \b)}
=   C_{5}^{(\a \b)}
&  \hskip 0.3in &
\t^{(\a \b)}  \in \{ \t^{(02)},\t^{(03)} \}
 \\
f_2^{-1} H_{-}^{(\a \b)}
=   C_2^{(\a \b)}
&  &
\t^{(\a \b)}  \in \{ \t^{(00)},\t^{(10)} ,\t^{(20)}\}
\no \\
f_1^{-1} H_{+}^{(\a \b)}
= C_1^{(\a \b)}
&  &
\t^{(\a \b)} \in \{  \t^{(10)},\t^{(12)},\t^{(13)},\t^{(20)},\t^{(22)} ,\t^{(23)}\}
\no
\eea
When $h=0$, the additional relations are,
\bea
\label{zeroh}
f_2^{-1} H_{-}^{(\a \b)}
=   C_{2}^{(\a \b)}
& \hskip 0.7in & \tau ^{( \a \b )} \in \{ \tau ^{(01)}, \tau ^{(11)}, \t^{(21)} \}
\no \\
f_1^{3} H_{-}^{(\a \b)}
= C_1^{(\a \b)}
&  &
\t^{(\a \b)} \in \{  \t^{(11)},\t^{(21)}\}
\eea
If both $h=g_9=0$, there are some additional relations besides all the above ones,
\bea
f_2^{-1} H_{-}^{(\a \b)}
=   C_{2}^{(\a \b)}
& \hskip 0.7in & \tau ^{( \a \b )} \in \{ \tau^{(02)}, \tau ^{(12)}, \tau^{(22)} \}
\no \\
f_1^{3} H_{-}^{(\a \b)}
= C_1^{(\a \b)}
&  &
\t^{(\a \b)} \in \{ \tau^{(02)}, \tau ^{(03)}\}
\eea
Note that, by linearity of the BPS equation, and the fact
that $f_5$ and $\psi^\dagger \psi + \xi ^\dagger \xi$ must be positive,
we may rescale both $\xi$ and $\psi$ by the same real
constant, and without loss of generality choose the following normalization,
\bea
\label{norm2}
H_{+}^{(00)} = f_5
\eea
In the next section, we shall show that similar unique correspondences
exist for $f_1$ and $f_2$.

\section{Vanishing Hermitian forms}
\setcounter{equation}{0}

In the preceding section, it was shown that certain Hermitian
forms of the supersymmetry spinors are related to the metric factors $f_1, f_2, f_5$,
in terms of unknown constants. In the present section, we shall show that  certain
Hermitian forms vanish automatically as a result of the BPS equations.
These relations will be obtained using the reality properties of various combinations.
Their derivations will involve combinations in which the quantities $p_z, p_{\bar z}^*,
g_z$, and $g_{\bar z}^*$ are algebraically eliminated, since these functions do
not, generically, obey definite reality conditions.
In this section, we shall use the following definition for Hermtian forms,
\bea
H_{\pm}^{(\a \b)} & \equiv &
\psi^{\dagger} \t^{(\a \b)} \psi \pm \xi^{\dagger} \t^{(\a \b)} \xi
\no \\
H_{g\pm}^{(\a \b)} & \equiv &
g_9 \psi^{\dagger}  \t^{(\a \b)} \xi \pm  g^*_9 \xi^{\dagger} \t^{(\a \b)} \psi
\no \\
H_{h\pm}^{(\a \b)} & \equiv &
h \psi^{\dagger}  \t^{(\a \b)} \xi \pm h^* \xi^{\dagger} \t^{(\a \b)} \psi
\eea
where the first line reproduces the definition already used in (\ref{defH}).
Note that $H_\pm ^{(\a \b)}$, $H_{g+} ^{(\a \b)}$, and $H_{h+} ^{(\a \b)}$
are real, while $H_{g-} ^{(\a \b)}$, and $H_{h-} ^{(\a \b)}$ are purely imaginary.

\subsection{Example of Hermitian relation calculation}
We leave the detailed calculation in appendix \ref{appVHF}, and quote the results below.
There are three different ways to compute vanishing hermitian forms, each will provide additional
relations among Hermitian forms, which are also summarized below.

Before giving the complete list of all Hermitian relations, we illustrate one example about computing Hermitian forms.

Considering a combination of the BPS equations of (\ref{BPS2a}),
given by 2(m) + (i) + (9). All $f, g_z ,
g_{\bar z}, h, h^*$ terms cancel in these combinations, and the resulting equations are,
\bea
& &
D_z \ln (f_5^2 f_2 f_1) \psi + \left({2i \over f_5} \t^{(22)} + {i\over f_2} \t^{(11)}
- {\nu \over f_1} \t^{(23)}\right) \xi^* - {i\over 2} g_9 \t^{(13)} \psi^* = 0
\no \\
&&
D_z \ln (f_5^2 f_2 f_1) \xi + \left(-{2i \over f_5} \t^{(22)} + {i\over f_2} \t^{(11)}
+ {\nu \over f_1} \t^{(23)}\right) \psi^* + {i\over 2} g^*_9 \t^{(13)} \xi^* = 0
\eea
Multiplying the first equation by $\xi^t $, the second by $-\psi^t $,
adding both to cancel the differential terms, and taking the transpose, we obtain,
\bea
0& =&  \xi^\dagger  \left({2i \over f_5} \t^{(22)} + {i\over f_2} \t^{(11)}
+ {\nu \over f_1} \t^{(23)}\right) \xi
- {i\over 2} g_9 \psi^\dagger \t^{(13)}  \xi
\no \\
&&
+\psi^\dagger  \left( {2i \over f_5} \t^{(22)} - {i\over f_2} \t^{(11)}
+ {\nu \over f_1} \t^{(23)}\right)  \psi
- {i\over 2} g^*_9 \xi^\dagger \t^{(13)}  \psi
\eea
Each term in $\xi ^\dagger \tau \xi$ or $\psi ^\dagger \tau \psi$, and each combination
of terms involving $g_9$ and $g_9^*$, is either real or purely imaginary.
Separating real and imaginary parts of the above equations, we have,
\bea
H_+^{(23)} =0
\hskip 0.5in
{2 \over f_5} H_+^{(22)}
- {1 \over f_2} H_-^{(11)}
- \half H_{g+}^{(13)} =0
\eea
So we obtained one vanishing Hermitian form, and a relation among them.

\subsection{Summary of all Hermitian relations}

As a summary, we list all the vanishing Hermitian relations below,
\bea
\label{Hermit1}
H_+^{(\a \b)} = 0 && (\a \b) \in \bigg \{ (10),(11),(13),(20),(21),(23),(30),(31),(32),(33) \bigg \}
\no \\
H_-^{(\a \b)} = 0 && (\a \b) \in \bigg \{ (00),(01),(02),(03),(12),(22) \bigg \}
\no \\ \no \\
H_{g+}^{(\a \b)}  = 0
&& (\a \b) \in \bigg \{ (00),(02),(03),(10),(12),(20),(22) \bigg \}
\no \\
H_{g-}^{(\a \b)}  = 0
&& (\a \b) \in \bigg \{ (01),(30),(32),(33) \bigg \}
\no \\ \no \\
H_{h+}^{(\a \b)}  = 0
&& (\a \b) \in \bigg \{ (00),(31),(32),(33) \bigg \}
\no \\
H_{h-}^{(\a \b)}  = 0
&& (\a \b) \in \bigg \{ (01), (02),(03),(12),(22) \bigg \}
\eea
The remaining equations from the first set are as follows,
\bea
\label{firstset}
(00) &&
{2 \over f_5} H_+^{(22)}
- {1 \over f_2} H_-^{(11)}
- \half H_{g+}^{(13)} =0
\no \\
(01) &&
 {1\over f_2} H_-^{(10)}
- {\nu \over f_1} H_+^{(22)}=0
\no \\
(02) &&
{2 \over f_5} H_-^{(20)}
- {\nu \over f_1} H_-^{(21)}
- { i \over 2} H_{g-}^{(11)} =0
\no \\
(12) &&
{2\over f_5} H_-^{(30)}
+{1\over f_2} H_+^{(03)}
-{\nu \over f_1} H_-^{(31)}=0
\no \\
(22) &&
{2 \over f_5} H_+^{(00)}
+ {1 \over f_2} H_-^{(33)}
- {\nu \over f_1} H_+^{(01)}
- \half  H_{g+}^{(31)} =0
\no \\
(30) &&
{2 \over f_5} H_+^{(12)}
+ {1 \over f_2} H_-^{(21)}
+ \half  H_{g+}^{(23)} =0
\no \\
(31) && {1\over f_2} H_-^{(20)}
+ {\nu \over f_1} H_+^{(12)}=0
\no \\
(32) &&
{2 \over f_5} H_-^{(10)}
- {\nu \over f_1} H_-^{(11)}
+ {i \over 2}  H_{g-}^{(21)} =0
\eea
Those of the second set are as follows,
\bea
\label{secondset}
(02)
&& {i \over f_5} H_-^{(20)}
- {1 \over 8} H_{g-}^{(11)}
- {i \over 8} H_{h+}^{(10)}=0
\no \\
(12) &&
{1\over f_5} H_-^{(30)}
 -  f H_+^{(00)}  = 0
\no \\
(32)
&&
{1 \over f_5} H_-^{(10)}
- {i \over 8} H_{g-}^{(21)}
+ {1 \over 8} H_{h+}^{(20)}=0
\no \\
(02)
&&  3 H_{g-}^{(11)}
-i H_{h+}^{(10)}=0
\no \\
(12) && {1\over f_2} H_+^{(03)}
 +  f H_+^{(00)} = 0
\no \\
(32)
&&
{3i} H_{g-}^{(21)}+  H_{h+}^{(20)} =0
\eea
Finally, those of the third set are given by,
\bea
\label{thirdset}
(00)
&& {1 \over  f_5}H_+^{(22)}
+ {1 \over 2} H_{g+}^{(13)} = 0
\no \\
(01)
&& {i\nu\over  f_1} H_{+}^{(22)}
- 2i f H_{-}^{(13)}
- {1 \over 2} H_{h-}^{(13)}=0
\no \\
(03)
&& 2i f H_{-}^{(11)}
+ {1 \over 2} H_{h-}^{(11)}=0
\no\\
(22) &&{i\over  f_5} H_{+}^{(00)}
+{i\nu \over  f_1}H_{+}^{(01)}
- 2if H_{-}^{(30)}
+{i \over 2}  H_{g+}^{(31)}
- {1 \over 2} H_{h-}^{(30)}=0
\no \\
(30) &&
{1\over  f_5} H_{+}^{(12)}
- {1 \over 2} H_{g+}^{(23)}= 0
\no\\
(31) &&
{\nu \over  f_1}H_{+}^{(12)}
+ 2 f H_{-}^{(23)}
- {i \over 2} H_{h-}^{(23)}=0
\no \\
(33)
&& 2 f  H_{-}^{(21)}
- {i \over 2} H_{h-}^{(21)}=0
\eea

\subsection{Implications for the metric factors}

The relations between the metric factors $f_5, f_2, f_1$ and Hermitian forms,
derived in (\ref{gen1}), together with the vanishing Hermitian forms of the
preceding section, produce vanishing value for the following constants,
\bea
0 & = &  C_5 ^{(31)} = C_5 ^{(32)} = C_5 ^{(33)} = C_5 ^{(01)}
\no \\
0 & = & C_2 ^{(31)} = C_2 ^{(32)}
\no \\
0 & = & C_1 ^{(30)} = C_1 ^{(32)} = C_1 ^{(33)}
\eea
The remaining non-vanishing Hermitian forms  $f_1, f_2, f_5$ are given as follows,
\bea
f_5^{-1} H_+^{(00)} & =  & 1
\hskip 1in
f_5^3 H_-^{(30)} =   c_5
\no \\
f_2^{-1} H_-^{(33)} & = &   c_2
\no \\
f_1^{-1} H_+^{(01)} & = & c_1
\eea
Here, we have used the normalization $C_5 ^{(00)}=1$ derived in (\ref{norm2}),
and abbreviated $C_1^{(01)}$, $C_2 ^{(33)}$, and $C_{5}^{(30)}$ by $c_1$, $c_2$,
and $c_5$ respectively, all three  of which are real constants.

\sm

A series of additional constant Hermitian forms were derived in (\ref{zerog})
and (\ref{zeroh}) respectively when $g_9=0$ and $h=0$ or both. These
relations are readily seen to follow directly from the equations
(\ref{firstset}), (\ref{secondset}), and (\ref{thirdset}), and may thus be omitted.

\section{General solution to the reduced BPS equations}

In this section, we shall exploit the vanishing hermitian forms to solve the BPS equations,
 and conclude that the only solution to the reduced BPS equations is $AdS_5 \times S^5$.

\subsection{Solving the  Hermitian relations $H_\pm ^{(\a \b)}=0$}

We begin by solving the Hermitian relations of the form $H_\pm ^{(\a \b)}=0$,
namely those relations given in the first two lines of (\ref{Hermit1}). Once we shall
have solved for those, we shall solve the remaining Hermitian equations of
(\ref{Hermit1}) as well as the remaining BPS equations.

\sm

Every set of four Hermitian relations can be grouped together to derive an equivalent
set of simplified relations between spinor components,
\bea
\label{group1}
H_-^{(00)} = H_-^{(03)} = H_+^{(30)} =H_+^{(33)} =0
& \qquad \Rightarrow \qquad &
\xi^*_{+\pm} \xi_{+\pm} - \psi^*_{-\pm} \psi_{-\pm} =0
\no \\ &&
\xi^*_{-\pm} \xi_{-\pm} - \psi^*_{+\pm} \psi_{+\pm} =0
\eea
\bea
\label{group2}
H_-^{(01)} = H_-^{(02)} = H_+^{(31)} =H_+^{(32)} =0
& \qquad \Rightarrow \qquad &
\xi^*_{+\pm} \xi_{+\mp} - \psi^*_{-\pm} \psi_{-\mp} =0
\no \\
&& \xi^*_{-\pm} \xi_{-\mp} - \psi^*_{+\pm} \psi_{+\mp} =0
\eea
\bea
\label{group3}
H_+^{(10)} = H_+^{(13)} = H_+^{(20)} =H_+^{(23)} =0
& \qquad \Rightarrow \qquad &
\xi^*_{\pm+} \xi_{\mp +} + \psi^*_{\pm+} \psi_{\mp +} =0
\no \\
&& \xi^*_{\pm -} \xi_{\mp -}+ \psi^*_{\pm -} \psi_{\mp -} =0
\eea
\bea
\label{group4}
H_+^{(11)} = H_-^{(12)} = H_+^{(21)} =H_-^{(22)} =0
& \qquad \Rightarrow \qquad &
\xi^*_{+\pm} \xi_{-\mp} + \psi^*_{+\mp} \psi_{-\pm} =0
\no \\
&& \xi^*_{-\pm} \xi_{+\mp} + \psi^*_{-\mp} \psi_{+\pm} =0
\eea
From the first group (\ref{group1}), we deduce that
$|\xi_{+ \pm}| = |\psi _{- \pm}|$,  $|\xi_{- \pm}| = |\psi_{+ \pm}|$.
Thus, the corresponding functions are related by phases, which we introduce
as follows,
\bea
\xi_{++} = e^{i \theta_1} \psi_{-+}
\hskip 1in
\xi_{+-} = e^{i \theta_2} \psi_{--}
\no \\
\xi_{-+} = e^{i \theta_3} \psi_{++}
\hskip 1in
\xi_{--} = e^{i \theta_4} \psi_{+-}
\eea
Substituting these results into the second group (\ref{group2}), we obtain,
\bea
\label{gr2}
\left ( e^{i(\theta_2 -\theta_1)} - 1 \right ) \psi^*_{-+} \psi_{--} & = & 0
\no \\
\left  ( e^{i(\theta_4 -\theta_3)} - 1 \right ) \psi^*_{++} \psi_{+-} & = & 0
\eea
while from the third and fourth groups (\ref{group3}) and (\ref{group4}),  we obtain,
\bea
\label{gr34}
e^{i(\theta_3 -\theta_1)} \psi^*_{-+} \psi_{++} + \psi^*_{++} \psi_{-+} & = & 0
\no \\
e^{i(\theta_4 -\theta_2)}  \psi^*_{--} \psi_{+-} + \psi^*_{+-} \psi_{--} & = & 0
\no \\
e^{i(\theta_4 -\theta_1)} \psi^*_{-+} \psi_{+-} + \psi^*_{+-} \psi_{-+} & = & 0
\no \\
e^{i(\theta_3 -\theta_2)}  \psi^*_{--} \psi_{++} + \psi^*_{++} \psi_{--} & = &0
\eea
The solution depends upon whether the spinor
components $\psi _{\pm \pm }$ generically vanish or not.

\subsubsection{The generic case for which
$\psi _{++} \psi _{+-} \psi _{-+} \psi _{--} \not= 0$}

When the components $\psi _{++}, \psi _{+-}, \psi _{-+}$, and $ \psi _{--} $
are all generic and non-vanishing, then equations (\ref{gr2})  require,
\bea
\label{theta1234}
\theta _1 = \theta _2 \hskip 1in \theta _3 = \theta _4
\eea
We introduce the angles $\Theta$ and $\Theta'$, defined by,
\bea
4 \Theta  & \equiv & \theta _1 - \theta _3 + \pi \hskip 1in
\theta _1 = + 2 \Theta + 2 \Theta'
\no \\
4 \Theta' & \equiv & \theta _1 + \theta _3 - \pi \hskip 1in
\theta _3 = - 2 \Theta + 2 \Theta' + \pi
\eea
Note that we also have
$4 \Theta = \theta _2 - \theta _3 + \pi
= \theta _1 - \theta _4 + \pi = \theta _2 - \theta _4 + \pi$
in view of (\ref{theta1234}).  The equations of (\ref{gr34}) amount to relations
between the phases of the components of $\psi$ which may all be expressed
in terms of the angle $\Theta$ as follows,
\bea
{\psi^*_{-+} \over \psi_{-+}} =  e^{4i \Theta}  {\psi^*_{++} \over \psi_{++}}
\hskip 1in
{\psi^*_{--} \over \psi_{--}} =  e^{4i \Theta}  {\psi^*_{+-} \over \psi_{+-}}
\no \\
{\psi^*_{-+} \over \psi_{-+}} =  e^{4i \Theta}  {\psi^*_{+-} \over \psi_{+-}}
\hskip 1in
{\psi^*_{--} \over \psi_{--}} =  e^{4i \Theta}  {\psi^*_{++} \over \psi_{++}}
\eea
or equivalently,
\bea
\label{equiv1}
{\psi^*_{++} \over \psi_{++}} = {\psi^*_{+-} \over \psi_{+-}}
\qquad \qquad
{\psi^*_{-+} \over \psi_{-+}} = {\psi^*_{--} \over \psi_{--}}
\qquad \qquad
{\psi^*_{-+} \over \psi_{-+}} = {(e^{- 2 i \Theta}\psi_{++})^* \over e^{- 2 i \Theta}\psi_{++}}
\eea
The solutions to these relations may be parametrized in terms of 4 real arbitrary
functions $r_{++}, r_{+-}, r_{-+}, r_{--}$, the angle $\Theta$, and a further arbitrary
angle $\Lambda$,
\bea
\label{psig}
\psi _{++} & = & r_{++} \, e^{ i \Lambda + i \Theta}
\no \\
\psi _{+-} & = &  r_{+-} \, e^{ i \Lambda + i \Theta}
\no \\
\psi _{-+} & = &  r_{-+} \, e^{ i \Lambda - i \Theta}
\no \\
\psi _{--} & = &  r_{--} \, e^{ i \Lambda - i \Theta}
\eea
The solutions for $\xi$ now follow immediately,
\bea
\label{xig}
\xi _ {++} & = & r_{-+} \, e^{ i \Lambda' + i \Theta }
\no \\
\xi _ {+-} & = & r_{--} \, e^{ i \Lambda' + i \Theta }
\no \\
\xi _ {-+} & = & - r_{++} \, e^{ i \Lambda' - i \Theta }
\no \\
\xi _ {--} & = & - r_{+-} \, e^{ i \Lambda' - i \Theta }
\eea
where $\Lambda ' = 2 \Theta' + \Lambda$. (Using the $U(1)$ gauge freedom,
we could remove the overall relative phase factor between $\psi$ and $\xi$,
namely $\Lambda ' - \Lambda$, and set this angle to zero so that
$\Lambda' = \Lambda$. To keep symmetry manifest, and to remain
as general as possible, we shall not make this gauge choice here.)
One is then left with 4 real functions
$r$, and two angles $\Theta$, and $\Lambda$.

\subsubsection{Non-generic cases with vanishing spinor components}

It is clear from the above generic solution that a non-generic solution could
be viewed as the limit of the generic case. For example, if $\psi _{--}=0$,
then this case can be viewed as the limit in which $r_{--} \to 0$. Tracking
this case back to the original equations, it is clear that one retains
the first and third equations of (\ref{equiv1}), whose solution simply
amounts to the generic solution in which we set $r_{--}=0$.

One can check that the only exceptions to the generic solution is when either
$\psi_{-+} = \psi_{--} =0, \psi_{++} \psi_{+-} \ne 0$ or
$\psi_{++} = \psi_{+-} =0, \psi_{-+} \psi_{--} \ne 0$, which has equivalent
derivations. We shall focus on the first possibility then, which may be parameterized
by four real functions $r_{++}, r_{+-}, \Lambda_1 , \Lambda_2$ , and equations (\ref{gr2}), (\ref{gr34}) are
automatically satisfied.
\bea
\label{psig1}
\theta_3 & = & \theta_4
\no \\
\psi _{++} & = & r_{++} \, e^{ i \Lambda_1} = r_1 \, e^{ i \Lambda_1}
\no \\
\psi _{+-} & = &  r_{+-} \, e^{ i \Lambda_2} = r_2 \, e^{ i \Lambda_2}
\no \\
\xi _{-+} & = &   r_1 \, e^{ i (\Lambda_1 + \theta_3)}
\no \\
\xi _{--} & = &   r_2 \, e^{ i (\Lambda_2 + \theta_3)}
\eea
We shall call this the second type of solution, the difference from the generic solution
being $\psi_{++}$ and $\psi_{+-}$ may have different phases.

\subsubsection{Two types of solutions}

The solutions of (\ref{psig}) and (\ref{xig}) constitute the first type of solution.
They may be expressed in matrix form, as follows,
\bea
\label{psixi}
\psi & = & e^ { i \Lambda + i \Theta \tau ^{(30)} } r
\no \\
\xi & = &  \left ( i \tau ^{(20)} \right )
e^ { i \Lambda ' - i \Theta \tau ^{(30)} }  \, r
\eea
where $r$ is given by,
\bea
r = \left ( \matrix{ r_{++} \cr r_{+-} \cr r_{-+} \cr r_{--} \cr} \right )
\equiv \left ( \matrix{ r_1 \cr r_2 \cr r_3 \cr r_4 \cr} \right )
\eea
We can readily check that this form of $\psi$ and $\xi$ indeed
reproduces all the vanishing Hermitian forms of the first two lines of (\ref{Hermit1}).
In checking this, the dependences on the angles $\Lambda$ and $\Lambda'$
clearly drop out,  but the dependence on $\Theta$ may or may not drop out. We find,
\bea
H^{(\a \b)} _\pm = r^t e^{ - i \Theta \tau^{(30)}} \left [
\tau ^{(\a \b)} \pm \tau ^{(20)} \left ( \tau ^{(\a \b)} \right )^t \tau ^{(20)} \right ]
e^{i \Theta \tau ^{(30)}} r
\eea
This combination will vanish for a given assignment of $(\a \b)$ and all values
of $r$ and $\Theta$ provided either
$(i)$ the terms in the square brackets cancel; or $(ii)$ $\tau^{(\a \b)}$
is anti-symmetric and commutes with $\tau ^{(30)}$ (which is the case
precisely only for $\tau^{(02)}$ and $\tau^{(32)}$). These two vanishing conditions
reproduce all the vanishing Hermitian forms of the first two lines of (\ref{Hermit1}),
as well as two additional relations,
\bea
H_+ ^{(02)} = H_- ^{(32)}=0
\eea
These relations do not figure amongst  the first two lines of (\ref{Hermit1}),
and remarkably never enter the additional relations either.
Equation (\ref{psig1}) gives the second type of solution, with non-vanishing components,
\bea
\label{psig2}
\psi _{++} & = &  r_1 \, e^{ i \Lambda_1}
\no \\
\psi _{+-} & = &  r_2 \, e^{ i \Lambda_2}
\no \\
\xi _{-+} & = &   r_1 \, e^{ i (\Lambda_1 + \theta_3)}
\no \\
\xi _{--} & = &   r_2 \, e^{ i (\Lambda_2 + \theta_3)}
\eea

\subsection{Solving the Hermitian relations $H_{g \pm}^{(\a \b)}$}

\subsubsection{For the first type of solution}

We shall focus on the first type of generic solution (\ref{psixi}) in the section.
Expressing $H_{g \pm} ^{(\a \b)}$ in terms of the above parametrization gives
\bea
H_{g \pm}^{(\a \b)} & = &
i g_9 \, e^{  i (\Lambda ' - \Lambda)} \, r^t e^{- i \Theta \tau^{(30)} }
\tau ^{(\a \b)} \tau ^{(20)}  e^{- i \Theta \tau ^{(30)}} r
\no \\ &&
\mp i g_9^* \, e^{- i (\Lambda ' - \Lambda)} \, r^t e^{ i \Theta \tau^{(30)} }
\tau ^{(20)} \tau ^{(\a \b)}   e^{ i \Theta \tau ^{(30)}} r
\eea
This form may be simplified according to whether $\tau ^{(\a \b)}$ commutes or
anti-commutes with $\tau ^{(30)}$, so that the formula may be split into two
cases,
\bea
\a = 0,3 & \hskip 1in &
H_{g \pm}^{(\a \b)}
=  i g_9 \, e^{  i (\Lambda ' - \Lambda)} \, r^t  \tau ^{(\a \b)} \tau ^{(20)}   r
\no \\ && \hskip 0.6in
\mp i g_9^* \, e^{- i (\Lambda ' - \Lambda)} \, r^t  \tau ^{(20)} \tau ^{(\a \b)}  r
\no \\
\a = 1,2 & \hskip 1in &
H_{g \pm}^{(\a \b)}
= i g_9 \, e^{  i (\Lambda ' - \Lambda)} \, r^t  e^{- 2 i \Theta \tau^{(30)} }
\tau ^{(\a \b)} \tau ^{(20)}   r
\no \\ && \hskip 0.6in
\mp i g_9^* \, e^{- i (\Lambda ' - \Lambda)} \, r^t  e^{2 i \Theta \tau^{(30)} }
\tau ^{(20)} \tau ^{(\a \b)}  r
\eea
It follows from the vanishing of $r^t \tau ^{(\gamma \delta)} r$, when $\tau ^{(\gamma \delta)}$
is anti-symmetric, that we have the following automatic cancelations,
\bea
H_{g+}^{(00)} = H_{g+}^{(03)} = H_{g+}^{(12)} = H_{g+}^{(22)}
= H_{g-}^{(01)} = H_{g-}^{(32)} =0
\eea
The remaining cases of the middle two lines of (\ref{Hermit1}), impose the following
conditions. For $H_{g-}^{(30)}$, $ H_{g-}^{(33)}$, and $H_{g+}^{(02)}$, we have,
\bea
\label{g9eq}
\left ( g_9 \, e^{  i (\Lambda ' - \Lambda)} - g_9^* \, e^{- i (\Lambda ' - \Lambda)} \right )
r ^t \tau ^{(\gamma \delta )} r =0
\hskip 0.6in
(\gamma \delta ) \in \bigg \{ (10), (13), (22) \bigg \}
\eea
while the vanishing of $H_{g+}^{(10)}$, and $ H_{g+}^{(20)}$,  imposes
the following relations,
\bea
\left ( \matrix{ i \sin (2 \Theta) \cG_- & - \cos ( 2 \Theta) \cG_+ \cr
i \cos (2 \Theta) \cG_- & \sin (2  \Theta) \cG_+ \cr } \right )
\left ( \matrix{ r^t r \cr r^t \tau ^{(30)} r \cr } \right ) =0
\eea
where $\cG_\pm \equiv g_9 \, e^{  i (\Lambda ' - \Lambda)} \pm g_9^* \, e^{- i (\Lambda ' - \Lambda)}$. Since $r^tr \not=0$, this equation must admit non-trivial solutions,
requiring that the determinant of the above $2\times 2$ matrix vanish, i.e. $\cG_+ \cG_-=0$.
It is manifest that $\cG_+=0$, with $\cG_- \not=0$, cannot be a viable solution. Hence, we
must have
\bea
\cG_- = g_9 \, e^{  i (\Lambda ' - \Lambda)} - g_9^* \, e^{- i (\Lambda ' - \Lambda)}=0
\eea
which then automatically solves also the equations (\ref{g9eq}).

Now if $g_9 \ne 0$, we have $\cG_+\ne 0$, and also the relation $r^t \tau ^{(30)} r =0 $.
\be
\label{tau30}
r_1^2 + r_2^2 = r_3^2 + r_4^2
\ee
\subsubsection{Vanishing of $g_9$ for first type of solution}
The above result contradicts the equation (12) in (\ref{secondset}), which implies,
\bea
 H_+^{(00)} = {1\over f f_5} H_-^{(30)}
=  {2\over f f_5} (r_1^2 + r_2^2 - r_3^2 - r_4^2) = 0
\eea
which is impossible since $ H_+^{(00)}$ is positive-definite, and proportional to $f_5$.
As a result, we may conclude that $g_9 = 0$.
Now, by rechecking equations (\ref{firstset}), (\ref{secondset}), (\ref{thirdset})
using the condition $g_9 = 0$, we may find some additional
vanishing Hermitian forms.
\bea
&& H_-^{(10)} = H_-^{(11)} =H_+^{(12)} H_-^{(20)} =H_-^{(21)} =H_+^{(22)} = 0
\no \\
&& H_{h+}^{(10)} = H_{h-}^{(11)} = H_{h+}^{(20)} =H_{h-}^{(21)} = 0
\eea
Combing with the original vanishing Hermitian forms, we obtain
\bea
\psi^{\dagger} \t^{(\a \b)} \psi = 0 & \qquad &
(\a\b) \in \{ 10,11,12,20,21,22\}
\eea
The (11),(12),(21),(22) equations are equivalent to the following ones,
\bea
\psi^*_{\pm +} \psi_{\mp -} = 0
\eea
from which we know that
\bea
\label{rr01}
r_1 r_4 =  r_2 r_3 = 0
\eea
From the (10) and (20) equations, we  also derive that
\bea
\label{rr02}
r_1 r_3 + r_2 r_4 = 0
\eea

\subsubsection{Vanishing of $g_9$ for second type of solution}
The computation of $H_{g \pm} ^{(\a \b)}$ equations is a lot simpler for
the second type of solution. Consider $H_{g +} ^{(10)}= H_{g +} ^{(20)} = 0$, using the
parametrization (\ref{psig2}), two equations combine to be,
\bea
g_9 ( r_1^2 + r^2_2 ) = 0
\eea
So we immediately have $g_9 = 0$.

\subsection{Non-existence of solutions except for $AdS_5 \times S^5$}

\subsubsection{First type of solution}

Without loss of generality, let's choose $r_4 = 0$ in equations (\ref{rr01}) and (\ref{rr02}).
It then follows that either $ r_1 = r_2 = 0 $ or $r_3 = 0$.
We now examine the dilatino equations in (\ref{BPS2a}),
\bea
(d_1) &\qquad&
4 i p_z  \xi
+  g_{z}  \psi
-  h  \tau^{(12)}  \xi^* = 0
\no \\
(d_2) &\qquad&
 4 i p^*_{\bar z}  \psi
+  g^*_{\bar z}  \xi
-  h^*  \tau^{(12)}  \psi^* = 0
\eea
If $r_1 = r_2 = r_4 = 0$, to have a non-vanishing value for $r_3$, we must have
\bea
p_z = p_{\bar z} = g_z = g_{\bar z} = h =0
\eea
From the analysis in section \ref{four}, we know that  this leads to the
$AdS_5 \times S^5$ solution.
On the other hand, if $r_3 = r_4 = 0$, the first dilatino equation reads in component form as,
\bea
0 &=& p_z \xi_{-\pm} =  p^*_{\bar z} \psi_{+\pm} 
\no \\
0 &=& g_z \psi_{++} + i h \psi_{+-}^* e^{-i\theta_3}
\qquad 
0 = i h^* \psi_{++} e^{i\theta_3} + g_z^* \psi_{+-}^*
\no \\
0 &=& g^*_{\bar z} e^{i\theta_3} \psi_{++} + i h^* \psi_{+-}^* 
\qquad
0 = i h \psi_{++} + g_{\bar z} e^{-i\theta_3} \psi_{+-}^*
\eea
In order to have non-vanishing spinor solutions, we should require
\bea
p_z = p_{\bar z}& = &  0
\no \\
|h|^2 + |g_z|^2  & = & 0
\no \\
|h|^2 + |g_{\bar z}|^2  & = & 0
\eea
The second dilatino equation provides similar results,
requiring $p_z = p_{\bar z} = g_z = g_{\bar z} = h = 0$.
The solution is still limited to $AdS_5 \times S^5$.

\subsubsection{Second type of solution}

The analysis is identical to the one used for the first type of solution, and we
again reach the conclusion that the only possible solution is $AdS_5 \times S^5$.

\newpage

\section{Ansatz with $SO(2,4) \times SO(4)$ symmetry}
\label{nine}

Another realization of the superalgebra $SU(2,2|2)$ in Type IIB supergravity 
is motivated by the system of a stack of D3-branes in the presence of probe 
D7-branes (see the Introduction). This realization requires extending the 
invariance by adjoining a purely bosonic $SU(2)$ group, so that we have full
$SU(2,2|2) \times SU(2)$ symmetry \cite{D'Hoker:2008ix}. Its bosonic 
symmetry is $SO(2,4) \times SO(4) \times SO(2)$. In this section, we 
shall use a less restrictive bosonic symmetry $SO(2,4) \times SO(4)$,
thereby increasing the generality of our results. We shall show
that the only solution is $AdS_5 \times S^5$ for this realization of the symmetry.

\subsection{Construction of the Ansatz}

We seek the most general Ansatz in Type IIB supergravity with
$ SO(2,4) \times SO(4)$ symmetry. The $SO(2,4)$-factor requires the 
geometry to contain an $AdS_5$-factor, while the  $SO(4)$-factor requires 
an $S^3$-factor, both of which are warped over a 2-dimensional surface 
$\Sigma$ with boundary. The total space-time then has the structure 
$AdS_5 \times S^3 \times \Sigma$.
Following a construction similar to the one given in section \ref{two}, the Ansatz 
for the metric is found to be,
\bea
ds^2 = f_5^2 ds^2 _{AdS_5} + f_3 ^2 ds^2 _{S^3} +  ds^2 _{\Sigma}
\eea
where $f_3$ and $f_5$  are  functions on $\Sigma$ only.
Orthonormal frames are defined by.
\bea
\label{frame2-1}
e^m & = & f_5 \, \hat e^m \hskip 1in m=0,1,2,3,4
\no \\
e^{i} & = & f_3 \, \hat e^{i} \hskip 1.1in i =5,6,7
\no \\
e^{a} \, &  & \hskip 1.4in a=8,9
\eea
Here, the metrics $ds^2 _{AdS_5}$ and $ds^2 _{S^3}$, as
well as the orthonormal frames $\hat e^m$ and $\hat e^{i}$,
refer to the spaces $AdS_5$ and $S^3$ with unit radius. In particular, we have
\bea
ds^2 _{AdS_5} & = & \eta _{mn} \, \hat e^m \otimes \hat e^n
\no \\
ds^2 _{S^3} & = & \delta _{i j} \, \hat e^{i} \otimes \hat e^{j}
\no \\
ds^2 _{\Sigma} & = & \delta _{ab} \, e^{a} \otimes e^{b}
\eea
The dilaton/axion field  $B$ is a function of $\Sigma$ only, 
and the associated composites $P,Q$ of (\ref{sugra1}) are decomposed as follows, 
$P= p_a e^a$, and $Q = q_a e^a$, where $p_a$ and $q_a$ are again functions
of $\Sigma$ only.
Finally, the most general Ans\"atze for the 3- and 5-form fields $G$ and $F_{(5)}$
consistent with $SO(2,4) \times SO(4)$ invariance, are given
by
\bea
F_{(5)} & = & f ( e^{01234} - e^{56789})
\no \\
G & = & i g e^{567}
\eea
By $SO(2,4) \times SO(4)$-invariance, the coefficient functions
$p_a, q_a, g$ and $f$ depend only on $\Sigma$. The functions
$f, q_a$ are real, while $p_a, g$ are complex-valued.

\subsection{Reducing the BPS equations}

The supersymmetry spinor $\ep$ must be covariant under $SO(2,4) \times SO(4)$, 
and may be built from the Killing spinors on the space $AdS_5 \times S^3$,
where each factor space has unit radius.

\subsubsection{Killing spinors}

We begin by constructing suitable Killing spinors on
$AdS_5 \times S^3$. We define a spinor representation
$\chi ^{\eta_1, \eta_2}$ of $SO(2,4) \times SO(4)$  by the following equations
\cite{D'Hoker:2007xy} \cite{D'Hoker:2008wc},
\bea
\label{KS-1}
\left ( \hat \nabla _m  - \half \eta_1 \gamma _m \otimes I_2  \right )
\chi ^{\eta _1, \eta _2}
    & = & 0 \hskip 1in m=0,1,2,3,4
\no \\
\left ( \hat \nabla _{i} - {i \over 2} \eta_2 I_4 \otimes \gamma _{i} \right )
\chi ^{\eta _1, \eta _2}
    & = & 0 \hskip 1in i=5,6,7
\eea
Here, $\g_m$ and $\g_i$ are the Dirac matrices on $AdS_5$ and $S^3$
respectively (see appendix \ref{appA2}).
Integrability  requires $\eta _1^2 = \eta _2 ^2 =1$. The
respective solution spaces are of dimension 4 and 2, so that
$\chi ^{\eta _1, \eta _2}$ has 8 independent solutions.
Charge conjugation $\chi \to \chi^c$ is given by,
\bea
\left ( \chi ^c \right ) ^{\eta _1, \eta _2}  = (B_{(1)} \otimes B_{(2)})^{-1}
\left ( \chi ^{\eta _1 , \eta _2} \right )^*
\eea
and reverses the sign of $\eta _1$, so that
$\left ( \chi ^c \right ) ^{\eta _1, \eta _2}$ is proportional to
$ \chi ^{-\eta _1 , \eta _2}$. The overall proportionality factor may be chosen freely;
here, we shall use the following convention,
\be
(\chi^c)^{\eta _1, \eta _2} = \chi ^{-\eta _1, \eta _2}
\ee
Putting all together, the full 32-component spinor $\ep$ may thus be decomposed
in terms of Killing spinors on $AdS_5 \times S^3$ as follows,
\bea
\label{ep1-1}
\ep = \sum _{\eta _1, \eta _2}
\chi ^{\eta _1, \eta _2}
\otimes
\l( \zeta _{\eta _1, \eta _2}
\otimes u_+  +  \zeta'_{\eta _1, \eta _2} \otimes u_-  \r)
\eea
We use the $u_\pm$ basis of (\ref{uu}), so that
$\zeta _{\eta _1, \eta _2}$ and $\zeta'_{\eta _1, \eta _2}$
are independent 2-dimensional spinor  functions on $\Sigma$.
The factor in parentheses in (\ref{ep1-1}) parametrizes a general 
4-component spinor. The 10-dimensional chirality condition $\G^{11} \ep = - \ep$, 
with   $\G^{11} = I_4 \otimes I_2 \otimes \s_3 \otimes \sigma _2$, 
reduces to $\zeta'_{\eta _1, \eta _2} = - i \s^3 \zeta_{\eta _1, \eta _2}$.
so that the spinor $\ep$ takes the following form,
\bea
\ep = \sum _{\eta _1, \eta _2}
\chi ^{\eta _1, \eta _2}
\otimes
\l( \zeta _{\eta _1, \eta _2}
\otimes u_+  -i \s^3 \zeta_{\eta _1, \eta _2} \otimes u_-  \r)
\eea
We shall continue to use the $\tau$ matrix notation as introduced in equation (\ref{tau}).

\subsubsection{The reduced BPS equations}

The detailed calculations for the reduction of the BPS equations
 are in Appendix C. Here, we summarize the final results. 
 The dilatino BPS equation is given by,
\bea
\label{dilatino1-1}
(d)  \qquad 0 =  g \zeta = p_a \s^a \s^1 \zeta^*
\eea
Any  solution with non-vanishing $\zeta$ requires $g=0$.
The second equation implies $\det(p_a \s^a) = 0$, from which we  derive
either $p_8 = i p_9 \ne 0$ or $p_8 = -i p_9 \ne 0$ or $p_8 = p_9 = 0$.
The gravitino equation decomposes into a system of 3 equations,
\bea
\label{gravitino1-1}
(m) &\qquad& 
0=({ 1\over 2 f_5}\t^{(30)} - {1\over 2} {D_{a} f_5 \over f_5} \s^{a} \s^3 - {f\over 2} )\zeta
\no\\
(i) &\qquad& 
0=({ 1\over 2 f_3}\t^{(03)} \s^3 - {1\over 2} {D_{a} f_3 \over f_3} \s^{a} \s^3 + {f\over 2})\zeta
\no \\
(a) &\qquad&
0 =  (D_{a} + {i\over 2}   \hat \omega_{a} \s^3 + {f \over 2} \s_a \s^3 - {i\over 2} q_a ) \zeta
\eea

\subsubsection{The chiral form of the reduced BPS equations}

Using the complex basis defined in (\ref{complex}),
we proceed to decompose the two-index spinor $\zeta$ into its two
chirality components in this same 2-dimensional spinor basis,
\be
\zeta_{\eta_1,\eta_2} = {\xi_{\eta_1,\eta_2} \choose \psi_{\eta_1,\eta_2}}
\ee
The reduced gravitino BPS equations now becomes
\bea
\label{BPS1-1}
(m_1)&\qquad&
\left ( {1 \over 2 f_5}\t^{(30)} - {f\over 2} \right ) \xi + {D_z f_5 \over 2 f_5} \psi = 0
\no \\
(m_2)&\qquad&
-{D_{\bar z} f_5 \over 2 f_5} \xi + \left ( {1 \over 2 f_5}\t^{(30)} - {f\over 2} \right ) \psi = 0
\no \\
(i_1)&\qquad&
({1 \over 2 f_3}\t^{(03)} + {f\over 2}) \xi + {D_z f_3 \over 2 f_3} \psi = 0
\no \\
(i_2)&\qquad&
-{D_{\bar z} f_3 \over 2 f_3} \xi + (-{1 \over 2 f_3}\t^{(03)} + {f\over 2}) \psi = 0
\no \\
(-_1)&\qquad&
(D_z + {i\over 2} \hat \omega_z - {i\over 2} q_z ) \xi= 0
\no \\
(-_2)&\qquad&
(D_z - {i\over 2} \hat \omega_z  - {i\over 2} q_z ) \psi + f \xi = 0
\no \\
(+_1)&\qquad&
(D_{\bar z} + {i\over 2} \hat \omega_{\bar z} - {i\over 2} q_{\bar z} ) \xi - f \psi= 0
\no \\
(+_2)&\qquad&
(D_{\bar z} - {i\over 2} \hat \omega_{\bar z}  - {i\over 2} q_{\bar z} ) \psi  = 0
\eea

\subsection{Solving the reduced BPS equations}

From the argument given below (\ref{dilatino1-1}), we have 
three possible solutions for $p_a$. We shall now show that only the $p_a=0$
solution is viable. Indeed, suppose that we had a solution with  $p_8 = i p_9 \ne 0$. From  
(\ref{dilatino1-1}), this implies that  $\psi = 0$. Then from $(-_2)$ in 
(\ref{BPS1-1}), we  find $f = 0$ in order to have non-zero $\xi$. 
But now the $(m_1)$ equation becomes ${1 \over 2 f_5}\t^{(30)} \xi = 0$, which 
in turn implies $\xi = 0$, making the resulting solution trivial.
The same arguments eliminates the case $p_8 = -i p_9 \ne 0$. Thus we are left with 
the only remaining solution $p_8 = p_9 = 0$. Now when $p_a = 0$, the Bianchi identity 
$dQ = 0$ implies that $q_a$ are constants. With the help of the
$U(1)_q$ gauge symmetry, we may  choose the gauge $q_a = 0$.

\subsubsection{Using discrete symmetries}

There are two commuting discrete symmetries for the reduced BPS equations,
\bea
T_1: \zeta \rightarrow \t^{(30)} \zeta
\no \\
T_2: \zeta \rightarrow \t^{(03)} \zeta
\eea
so we can project the solution to the eigenspace of $\t^{(30)}$ and $\t^{(03)}$,
\bea
\t^{(30)} \zeta = \nu \zeta
\no \\
\t^{(03)} \zeta = \gamma \zeta
\eea
It turns out that different projections will give the same solution for supergravity fields.
We restrict to the projection with $\nu = \gamma =1 $, so that only the $\zeta_{++}$ 
component can be non-zero; we shall abbreviate this component simply as $\zeta$.

\subsubsection{Hermitian forms for metric factors}

The calculations of Hermitian forms for metric factors proceed in
the same way as for $SO(2,4) \times SO(3) \times SO(2)$ case, and we 
will omit some details in the calculation.
Using differential equations $(-) (+)$ in equation (\ref{BPS1-1}), we find,
\bea
D_z ( \bar \psi \psi + \bar \xi \xi)  = 0
\eea
So $\bar \psi \psi + \bar \xi \xi$ is a constant, which may be normalized to $f_5$,
\bea
\label{f5-1}
f_5= \bar \psi \psi + \bar \xi \xi
\eea
since $f_5$ is a constant too, which is derived from $(m)$ equation in the next subsection.
Using $(+)$ and $ (-)$ and $(i)$ equations, we find 
$D_z (f_3^{-1} (\bar \psi \psi - \bar \xi \xi))=0$,
whose general solution is given by, $f_3 = c_3 (\bar \psi \psi - \bar \xi \xi)$.
Using the fact that $f f_5 =1 $ and the $(i)$ equation, the normalization factor $c_3$ is determined to be 1. So the expression for $f_3$ becomes,
\bea
f_3 = \bar \psi \psi - \bar \xi \xi
\eea

\subsubsection{Generic Solution to reduced BPS equations}

From equations $(m_1)$ and $(m_2)$, we have,
\bea
\left ( {1 \over 2 f_5}- {f\over 2} \right )^2 \psi  + \left  |{D_z f_5 \over 2 f_5} \right |^2 \psi =0
\eea
the solution to which is,
\bea
D_z f_5 =  0
\hskip 1in
f f_5 = 1
\eea
which imply that $f$ and $f_5$ are constants.
The solution to $(-_1)$ and $ (+_2)$ is given by,
\bea
\xi = \rho^{1/2} \overline{ \alpha(z)}
\no \\
\psi = \rho^{1/2}  \beta(z)
\eea
Where $\alpha(z), \beta(z)$ are holomorphic functions.
Using the above expression, the $(-_2)$ and $(+_1)$ equations now become,
\bea
(-_2)&\qquad&
\p_z (\rho \beta) + f \rho^2 \bar \alpha = 0
\no \\
(+_1)&\qquad&
\p_z (\rho \alpha) - f \rho^2 \bar \beta = 0
\eea
From the Hermitian form for $f_5$ in equation (\ref{f5-1}), we derive 
$\rho = f_5 /( |\alpha|^2+ |\beta|^2)$, and the differential equations become
\bea
(-_2)&\qquad&
\p_z \left ( {\beta\over |\alpha|^2+ |\beta|^2} \right ) 
+  {\bar \alpha\over (|\alpha|^2+ |\beta|^2)^2} = 0
\no \\
(+_1)&\qquad&
\p_z \left ( {\alpha \over |\alpha|^2+ |\beta|^2} \right ) 
- { \bar \beta \over (|\alpha|^2+ |\beta|^2)^2}= 0
\eea
These equations may be compared to equations (\ref{BPS-AdS5+}), 
which differ only by a factor
of $i$. Let $A(z)$ be an arbitrary holomorphic function, the solution is
\bea
\a (z) = { 1 \over \sqrt{\p_z A(z)}} \hskip 1in \beta (z) = -{  A(z) \over \sqrt{\p_z A(z)}}
\eea

\subsubsection{Uniqueness of the $AdS_5 \times S^5$}

The above solution coincides with $AdS_5 \times S^5$. This may be seen by
a convenient choice of the coordinate $z$, for which $A(z)=(e^z - ie^{-z})/(e^z + ie^{-z})$,
for which the metric manifestly factorizes and agrees with the unit radius metric on
$AdS_5 \times S^5$. 
Because there are 8 solutions to the Killing spinor $\chi^{\eta_1\eta_2}$,
as well as 4 solutions for spinor $\zeta$ from the different projections,
we recover indeed 32  supersymmetries.

\appendix

\newpage

\section{Clifford algebras}
\setcounter{equation}{0}


\subsection{Clifford algebra basis for $AdS_5 \times S^2 \times \Sigma   \times S^1$ space}
\label{appA1}
We choose a basis for the Clifford algebra which is well-adapted to the
$AdS_5 \times S^2 \times \Sigma   \times S^1$, with the frame
labeled as in (\ref{frame2}),
\bea
\G^m & = & \g^m \otimes I_2 \otimes I_2 \otimes \s^3 \hskip 1.1in m =0,1,2,3,4
\no \\
\G^{i} & = & I_4 \otimes \g^{i} \otimes I_2 \otimes \s^1 \hskip 1in i=5,6
\no \\
\G^{a} & = & I_4 \otimes \s^3 \otimes \g^{a} \otimes \s^1 \hskip 1in a = 7,8
\no \\
\G^{9} & = & I_4 \otimes \s^3 \otimes \g^{9} \otimes \s^1
\eea
where  a convenient basis for the lower dimensional Clifford algebras is as follows,
\bea
i \g^0 = \sigma ^2 \otimes I_2 & \hskip 1in & \g^5 = \g^7 = \sigma ^1
\no \\
\g^1 = \sigma ^1 \otimes I_2 & \hskip 1in & \g^6 = \g^8 = \sigma ^2
\no \\
\g^2 = \sigma ^3 \otimes \sigma ^2 & \hskip 1in & \g^9 = \sigma ^3
\no \\
\g^3 = \sigma ^3 \otimes \sigma ^1& \hskip 1in &
\no \\
\g^4 = \sigma ^3 \otimes \sigma ^3& \hskip 1in &
\eea

We also find the following matrix
\bea
\g^{01234}= i I_4 & \hskip 1in & \G^{01234} = i I_4 \otimes I_2 \otimes I_2 \otimes \s_3
\no \\
\g_{(2)}= -i\g^{56}=  \s^3 & \hskip 1in & \G_{(2)} = -i \G^{56} =  I_4 \otimes \s^3 \otimes I_2 \otimes I_2
\no \\
\g_{(3)}= -i\g^{78}=  \s^3 & \hskip 1in & \G_{(3)} = -i \G^{78} =  I_4 \otimes I_2 \otimes \s^3 \otimes I_2
\no \\
\eea
The 10-dimensional chirality matrix in this basis is given by
\bea
\G^{11} = \G^{0123456789}   = I_4 \otimes I_2 \otimes I_2 \otimes \s_2
\eea
The complex conjugation matrices in each subspace are defined by
\bea
\left ( \g^m \right ) ^* = - B_{(1)} \g ^m B_{(1)} ^{-1}
& \hskip .5in &   (B_{(1)})^* B_{(1)} = - I_2 \hskip .6in B_{(1)} = \s^3 \otimes \s^2
\no \\
\left ( \g^{i} \right ) ^* = + B_{(2)} \g ^{i} B_{(2)} ^{-1}
& &   (B_{(2)})^* B_{(2)} = + I_2 \hskip .6in B_{(2)} =  \g^5 = \s^1
\no \\
\left ( \g^{a} \right ) ^* = - B_{(3)} \g ^{a} B_{(3)} ^{-1}
& &   (B_{(3)})^* B_{(3)} = - I_2 \hskip .6in B_{(3)} =  \g^8 = \s^2
\no\\
\left ( \g^{9} \right ) ^* = - B_{(4)} \g ^{9} B_{(4)} ^{-1}
& &   (B_{(4)})^* B_{(4)} = - I_2 \hskip .6in B_{(4)} =  \g^8 = \s^2
\eea
where in the last column we have also listed the form of these matrices
in our particular basis. The 10-dimensional complex conjugation matrix $\cB$
is defined by $(\G^M)^* = \cB \G^M \cB^{-1} $ and $\cB \cB^* = I$, and in this
basis is given by
\bea
\cB =  B_{(1)} \otimes B_{(2)} \otimes B_{(3)} \otimes \s^1
    = \s^3 \otimes \s^2 \otimes \s^1 \otimes \s^2 \otimes \s^1
\eea

\subsection{Clifford algebra basis for $AdS_5 \times S^3 \times \Sigma$ space}
\label{appA2}

We choose another basis of the Clifford algebra for this ansatz, which is well-adapted to the
$AdS_5 \times S^3 $, with the frame
labeled as in (\ref{frame2-1}),
\bea
\G^m & = & \g^m \otimes I_2 \otimes I_2 \otimes \s^3 \hskip 1.1in m =0,1,2,3,4
\no \\
\G^{i} & = & I_4 \otimes \g^{i} \otimes I_2 \otimes \s^1 \hskip 1in i=5,6,7
\no \\
\G^{a} & = & I_4 \otimes I_2 \otimes \g^{a} \otimes \s^2 \hskip 1in a = 8,9
\eea
where  a convenient basis for the lower dimensional Clifford algebras is as follows,
\bea
\label{gamma-1}
i \g^0 = \sigma ^2 \otimes I_2 & \hskip 1in & \g^5 = \g^8 = \sigma ^1
\no \\
\g^1 = \sigma ^1 \otimes I_2 & \hskip 1in & \g^6 = \g^9 = \sigma ^2
\no \\
\g^2 = \sigma ^3 \otimes \sigma ^2 & \hskip 1in & \g^7 = \sigma ^3
\no \\
\g^3 = \sigma ^3 \otimes \sigma ^1& \hskip 1in &
\no \\
\g^4 = \sigma ^3 \otimes \sigma ^3& \hskip 1in &
\eea

We also find the following matrix
\bea
\g^{01234}= i I_4 & \hskip 1in & \G^{01234} = i I_4 \otimes I_2 \otimes I_2 \otimes \s_3
\no \\
\g_{\Sigma}= -i\g^{89}=  \s^3 & \hskip 1in & \G_{\Sigma} = -i \G^{89} =  I_4 \otimes I_2 \otimes \s_3 \otimes I_2
\eea
The 10-dimensional chirality matrix in this basis is given by
\bea
\G^{11} = \G^{0123456789}   = I_4 \otimes I_2 \otimes \s_3 \otimes \s_2
\eea
The complex conjugation matrices in each subspace are defined by
\bea
\left ( \g^m \right ) ^* = - B_{(1)} \g ^m B_{(1)} ^{-1}
& \hskip .5in &   (B_{(1)})^* B_{(1)} = - I_2 \hskip .6in B_{(1)} = \s^3 \otimes \s^2
\no \\
\left ( \g^{i} \right ) ^* = - B_{(2)} \g ^{i} B_{(2)} ^{-1}
& &   (B_{(2)})^* B_{(2)} = - I_2 \hskip .6in B_{(2)} =  \g^6 = \s^2
\no \\
\left ( \g^{a} \right ) ^* = - B_{(3)} \g ^{a} B_{(3)} ^{-1}
& &   (B_{(3)})^* B_{(3)} = - I_2 \hskip .6in B_{(3)} =  \g^9 = \s^2
\eea
where in the last column we have also listed the form of these matrices
in our particular basis. The 10-dimensional complex conjugation matrix $\cB$
is defined by $(\G^M)^* = \cB \G^M \cB^{-1} $ and $\cB \cB^* = I$, and in this
basis is given by
\bea
\cB =  B_{(1)} \otimes B_{(2)} \otimes B_{(3)} \otimes \s^2
    = \s^3 \otimes \s^2 \otimes \s^2 \otimes \s^2 \otimes \s^2
\eea
\newpage

\section{Reducing the BPS equations: $AdS_5 \times S^2 \times \Sigma \times S^1$ case}
\setcounter{equation}{0}
\label{appBPS}

In this appendix, we shall present the detailed calculations required for the
reduction of the BPS equations in $AdS_5 \times S^2 \times \Sigma \times S^1$ case.
The reduction of the BPS equations (\ref{BPS}) to our Ansatz  requires
the following combinations,
\bea
P \cdot \G & = & p_a \G ^a
\no \\
G \cdot \G & = & 6 \, \left ( i  g_a \G^{56a} + i g_9 \G^{569} + h \G^{789} \right )
\no \\
F_{(5)} \cdot \G & = & 240\, f \, \G^{01234}
\eea
The corresponding $\G$-matrices, in the basis given in Appendix \ref{appA1},
are given by,
\bea
\G^{a} & = & I_4 \otimes \s^3 \otimes \g^a \otimes \s^1
\no \\
\G^{56\tilde a} & = & i I_4 \otimes I_2 \otimes \g^{\tilde a} \otimes \s^1
\no \\
\G^{789} & = & i I_4 \otimes \s^3 \otimes I_2 \otimes \s^1
\no\\
\G^{01234} & = & i I_4 \otimes I_2 \otimes I_2 \otimes \s^3
\eea
We shall also need the decompositions of  $\cB ^{-1} \ep ^*$,
\bea
\cB^{-1} \varepsilon^* = \sum _{\eta _1, \eta _2,\eta_3}
\chi ^{\eta _1, \eta _2} \chi^{\eta_3} \otimes
* \zeta _{\eta _1, \eta _2,\eta_3} \otimes \theta
\eea
where we use the abbreviation,
\bea
\label{star}
* \zeta _{\eta _1, \eta _2,\eta_3}
& = &  -  \sigma^2 \eta _2 \zeta _{-\eta _1, -\eta _2,-\eta_3 }^*
\no \\
{}* \zeta & = & - i \tau ^{(121)}  \sigma ^2 \zeta ^* \eea in
$\tau$-matrix notation.

\subsection{The dilatino equation}

The dilatino equation is,
\bea
0=  i (P \cdot  \G) \cB^{-1} \ep ^* -{i\over 24} \G \cdot  G \ep
\eea
Using the above form of the $\G$-matrices, we obtain,
\be
0  =   \sum_{\eta_1 ,\eta_2,\eta_3 } \chi^{\eta_1 ,\eta_2 } \chi^{\eta_3}
\otimes
\left[ i p_a \s^a \sigma^2  \eta _2 \zeta _{-\eta _1, \eta _2,-\eta_3 }^*
+ {1\over 4} ig_{\tilde a } \s^{\tilde a } \zeta_{\eta _1, \eta _2 ,\eta_3}
+ {1\over 4} h \zeta_{\eta _1, - \eta _2,\eta_3 } \right]\otimes \theta^*
\ee
Since the $\chi^{\eta_1, \eta_2} \chi ^{\eta _3}$ are linearly independent,
each coefficient is required to vanish. The result may be recast economically 
using the $\tau$-matrix notation, and upon multiplication on the left by 
$-4i \tau^{(131)}$, we have,
\bea
0 = 4 p_a  \sigma^a \sigma^2 \zeta^*
+  g_{\tilde a} \tau^{(131)}  \sigma^{{\tilde a}} \zeta
+  h  \tau^{(121)}  \zeta
\eea

\subsection{The gravitino equation}

The gravitino equation is
\bea
\label{gravitino}
0 & = & d \ep + \omega \ep + \phi \ep + g \cB^{-1} \ep^*
\no \\
\omega &=& {1 \over 4} \omega_{AB} \Gamma^{AB} \no\\
\phi &=& -{i \over 2}Q + {i \over 480} (\Gamma \cdot F_{(5)}) e_A \G^A \no\\
g &=& - {1 \over 96} e_A \bigg( \G^A (\G \cdot G) + 2 (\G \cdot G) \G^A \bigg)
\eea
The non-vanishing spin connection components are as follows,
\bea
\label{spincon}
\omega^m {}_n = \hat \omega^m {}_n
&\qquad&
\omega^m {}_{a} = e^m {D_{a} f_5 \over f_5}
\no\\
\omega^{i} {}_{j} = \hat \omega^{i} {}_{j}
&\qquad&
\omega^{i} {}_{a} = e^{i} {D_{a} f_2 \over f_2}
\no\\
\omega^{a} {}_{b}
&\qquad&
\omega^{9} {}_{a} = e^{9} {D_{a} f_1 \over f_1}
\eea
where $D_a = \rho^{-1} \p_a$.
The hats refers to the canonical connections on $AdS_5$, $S^2$ respectively.
Projecting the spin-connection along the various directions we have the expression for $d\ep + \omega \ep$
\bea
(m) &\qquad& \nabla_m^\prime \ep + \half {D_{a} f_5 \over f_5} \G_{m} \G^{a} \ep
\no\\
(i) &\qquad& \nabla_{i}^\prime \ep + \half {D_{a} f_2 \over f_2} \G_{i} \G^{a} \ep
\no\\
(a) &\qquad& \nabla_{a} \ep
\no\\
(9) &\qquad& \nabla_{9}^\prime \ep + \half {D_{a} f_1 \over f_1} \G_{9} \G^{a} \ep
\eea
where the prime on the covariant derivative indicates that only the connection
along $AdS_5$, $S^2$ respectively is included.
Using the Killing spinor equations (\ref{KS}) we eliminate the primed covariant derivatives,
and bring out overall factors of $\G_M$, 
which yields
\bea
(m) &\qquad&  \G_{m}
\sum_{\eta_1 ,\eta_2,\eta_3 }  \chi^{\eta_1, \eta_2} \chi^{\eta_3}\otimes
\left[ {i \eta_1\over 2 f_5} \zeta_{\eta_1, \eta_2,\eta_3} + \half {D_{a} f_5 \over f_5} \s^{a} \zeta_{\eta_1, -\eta_2,\eta_3} \right ] \otimes  \theta^*~
\no\\
(i) &\qquad&  \G_{i}
\sum_{\eta_1 ,\eta_2,\eta_3 }  \chi^{\eta_1, \eta_2} \chi^{\eta_3}\otimes
\left[ {i \eta_2\over 2 f_2} \zeta_{\eta_1, \eta_2,\eta_3} + \half {D_{a} f_2 \over f_2} \s^{a} \zeta_{\eta_1, -\eta_2,\eta_3} \right ] \otimes  \theta^*~
\no\\
(a)  &\qquad& \nabla_{a} \ep
\no\\
(9)&\qquad&  \G_{9}
\sum_{\eta_1 ,\eta_2,\eta_3 }  \chi^{\eta_1, \eta_2} \chi^{\eta_3} \otimes
\left[ {i \eta_3\over 2 f_1} \s^3\zeta_{\eta_1, -\eta_2,\eta_3} +\half {D_{a} f_1 \over f_1} \s^{a} \zeta_{\eta_1, -\eta_2,\eta_3} \right ] \otimes  \theta^*
\eea
All terms in the gravitino equations (m), (i), and $(9)$, will contain 
$\G_M \chi^{\eta_1, \eta_2 }\chi_{\eta_3}$,
and we must require the coefficients to vanish independently.  The coefficients of
$\G_M \chi^{\eta_1 ,\eta_2} \chi^{\eta_3}$ can be expressed in the $\tau$-matrix
notation as
\bea
\label{redomega2}
(m) &\qquad& {i \over 2 f_5} \tau^{(300)} \zeta + \half {D_{a} f_5 \over f_5} \tau^{(010)} \s^{a} \zeta
\no\\
(i) &\qquad&  {i \over 2 f_2}  \tau^{(030)}\zeta + \half {D_{a} f_2 \over f_2} \tau^{(010)} \s^{a} \zeta
\no\\
(9) &\qquad&  {i \over 2 f_1}  \tau^{(013)}\s^3\zeta + \half {D_{a} f_1 \over f_1} \tau^{(010)} \s^{a} \zeta
\eea
 The $F_{(5)}$ part in the calculation of $\phi$ is as follows, 
\bea
{i \over 480} (\G \cdot F_{(5)}) \G^A e_A \ep
=  {i \over 2} f \G^{01234} \G^A e_A \ep
\eea
Projecting along the various directions, we have
\bea
(m) &\qquad& \G_m \sum_{\eta_1, \eta_2,\eta_3} \chi^{\eta_1, \eta_2} \chi^{\eta_3}
\otimes -{i \over 2}  f  \zeta_{\eta_1,  \eta_2,\eta_3}\otimes \theta^*~
\no\\
(i) &\qquad& \G_i \sum_{\eta_1, \eta_2,\eta_3} \chi^{\eta_1, \eta_2} \chi^{\eta_3}
\otimes {i \over 2}  f  \zeta_{\eta_1,  \eta_2,\eta_3}\otimes \theta^*~
\no\\
(a) &\qquad& \sum_{\eta_1, \eta_2,\eta_3} \chi^{\eta_1, \eta_2} \chi^{\eta_3}
\otimes \left[ - {i q_{a} \over 2}  \zeta_{\eta_1,  \eta_2,\eta_3} + {i \over 2}  f \s^{a} \zeta_{\eta_1, - \eta_2,\eta_3} \right] \otimes \theta~
\no\\
(9) &\qquad& \G_{9} \sum_{\eta_1, \eta_2,\eta_3} \chi^{\eta_1, \eta_2} \chi^{\eta_3}
\otimes \left[ {i \over 2}  f  \zeta_{\eta_1,  \eta_2,\eta_3} \right] \otimes \theta^*
\eea
 Using the $\tau$-matrix notation,  we write the coefficients of
$\G_M \chi^{\eta_1, \eta_2 }  \chi^{\eta_3}$ or $\chi^{\eta_1, \eta_2 } \chi^{\eta_3}$ in the form
\bea
\label{redphi2}
(m) &\qquad& -{i \over 2}  f  \zeta \no \\
(i) &\qquad& {i \over 2}  f  \zeta\no\\
(a) &\qquad& - {i q_{a} \over 2}    \zeta + {i f\over 2}\tau^{(010)} \s^{a}   \zeta \no\\
(9)&\qquad&  {i \over 2}  f  \zeta
\eea
The relevant expression for the calculation of $g$  is as follows,
\bea
g \cB^{-1} \epsilon^* = - {3! \over 96} e_A \bigg(ig_{\tilde  a} (\G^A \G^{56{\tilde a}} + 2 \G^{56{\tilde  a}} \G^A)
+  h(\G^A \G^{789} + 2 \G^{789} \G^A) \bigg) \cB^{-1} \epsilon^*
\eea
A few useful equations are as follows,
\bea
\G^{\tilde a} \G^{56\tilde b} + 2 \G^{56\tilde b} \G^{\tilde a} &=& \G^{56} (3 \delta^{\tilde a\tilde b} - \G^{\tilde a\tilde b})
= i \gamma_{(2)} (3 \delta^{\tilde a\tilde b} -  \sigma^{\tilde a\tilde b})
\no\\
\G^{\tilde a} \G^{789} + 2 \G^{789} \G^{\tilde a} &=& 3\G^{\tilde a} \G^{789}
\eea
Projecting along the various directions we obtain
\bea
(m) &\quad&  \G_m \sum_{\eta_1, \eta_2,\eta_3} \chi^{\eta_1 ,\eta_2 } \chi^{\eta_3}
\otimes {1 \over 16} ( - g_{\tilde a} \sigma^{{\tilde a}} * \zeta_{ \eta_1, \eta_2,\eta_3 }
+ i h  * \zeta_{ \eta_1, -\eta_2,\eta_3})\otimes \theta^*
\\
(i) &\quad&  \G_i \sum_{\eta_1, \eta_2,\eta_3} \chi^{\eta_1 ,\eta_2 } \chi^{\eta_3}
\otimes {1 \over 16} ( 3 g_{\tilde a} \sigma^{{\tilde a}} * \zeta_{ \eta_1, \eta_2,\eta_3 }
+ i h  * \zeta_{ \eta_1, -\eta_2,\eta_3})\otimes \theta^*
 \no\\
(a)&\quad&   \sum_{\eta_1, \eta_2,\eta_3} \chi^{\eta_1 ,\eta_2 } \chi^{\eta_3}
\otimes {1 \over 16} ( (3 g^{{a}} - g_{\tilde b} \sigma^{{a}{\tilde b}} ) * \zeta_{ \eta_1, -\eta_2,\eta_3 }
- 3i h  \sigma^{a} * \zeta_{ \eta_1, \eta_2,\eta_3})\otimes \theta
 \no\\
(9)&\quad&  \G_{9} \sum_{\eta_1, \eta_2,\eta_3} \chi^{\eta_1 ,\eta_2 } \chi^{\eta_3}
\otimes {1 \over 16} ( (3 g^{9} \sigma^{9}-  g_a \s^a) * \zeta_{ \eta_1, \eta_2,\eta_3 }
- 3i h  * \zeta_{ \eta_1, -\eta_2,\eta_3})\otimes \theta^*
\no
\eea
Using the $\tau$-matrix notation, we write the coefficient of
$\G_M \chi^{\eta_1, \eta_2} \chi^{\eta_3}$ or $\chi^{\eta_1, \eta_2}  \chi^{\eta_3}$ in the form
\bea
\label{redg2}
(m) &&
{1 \over 16} ( - g_{\tilde a} \sigma^{\tilde a} * \zeta
+ i h \tau^{(010)} * \zeta)
\\
(i) &&
{1 \over 16} ( 3 g_{\tilde a} \sigma^{\tilde a}  * \zeta
+ i h \tau^{(010)} * \zeta)
\no\\
(a) &&
{1 \over 16} ( 3 g_a \tau^{(010)}  * \zeta - g_{\tilde b} \tau^{(010)}\s_a^{~\tilde b} * \zeta
- 3i h  \sigma_a*  \zeta)
\no\\
(9) &&
{1 \over 16} ( 3 g^{9} \sigma^{3}  * \zeta - g_a \s^a  * \zeta
- 3i h \tau^{(010)} * \zeta) \no
\eea

\subsubsection{Assembling the complete gravitino BPS equation}

Now we combine the three equations $(\ref{redomega2})$, $(\ref{redphi2})$,
and $(\ref{redg2})$ to obtain the reduced gravitino equations.
We again argue that the $\Gamma_M \chi^{\eta_1, \eta_2}$ or $\chi^{\eta_1, \eta_2}$
are linearly independent which leads to the equations (\ref{gravitino1}).

For equations (m), (i) and $(9)$, we have dropped an overall factor of $\G_M$.
In equation $(a)$, $\hat \omega _{a} = (\hat \omega _{78})_{a}$ is the spin connection along $\Sigma$,
and we have used the connection formula $(\ref{spincon})$
for the covariant derivative and the fact $\gamma^{78} = i \, \sigma^3$.

\section{Reducing the BPS equations: $AdS_5 \times S^3 \times \Sigma$ case}
\setcounter{equation}{0}
\label{appC}

In this appendix, we shall present the detailed calculations required for the
reduction of the BPS equations in $AdS_5 \times S^3 \times \Sigma$ case.
The corresponding $\G$-matrices, in the basis given in appendix \ref{appA2},
are given by,
\bea
\G^{a} & = & I_4 \otimes I_2 \otimes \g^a \otimes \s^2
\no \\
\G^{567} & = & i I_4 \otimes I_2 \otimes I_2 \otimes \s^1
\no \\
\G^{01234} & = & i I_4 \otimes I_2 \otimes I_2 \otimes \s^3
\eea
We shall also need the decompositions of  $\cB ^{-1} \ep ^*$,
\bea
\cB^{-1} \varepsilon^* = \sum _{\eta _1, \eta _2}
\chi ^{\eta _1, \eta _2}  \otimes
(-i \s^1 \t^{(10)}\zeta^*_{\eta_1,\eta_2} \otimes u_+ + i \s^2 \t^{(10)}\zeta^*_{\eta_1, \eta_2} \otimes u_-)
\eea

\subsection{The dilatino equation}

The dilatino equation becomes,
\bea
0 & = & i p_a \G^a \sum _{\eta _1, \eta _2}
\chi ^{\eta _1, \eta _2}  \otimes
(-i \s^1 \zeta^*_{-\eta_1,\eta_2} \otimes u_+ + i \s^2 \zeta^*_{-\eta_1, \eta_2}\otimes  u_-)
\no \\
&&
+ {1\over 4} g \G^{567}
 \sum _{\eta _1, \eta _2}
\chi ^{\eta _1, \eta _2}
\otimes
\l( \zeta _{\eta _1, \eta _2}
\otimes u_+  -i \s^3 \zeta_{\eta _1, \eta _2} \otimes u_-  \r)
\eea
Using the above form of the matrices, we obtain,
\bea
0 &=& i p_a \s^a \sigma^2  \zeta _{-\eta _1, \eta _2}^*
+ {1\over 4} g\s^3 \zeta_{\eta _1, \eta _2}
\no \\
0 & = &  i  p_a \s^a \s^1 \zeta^*_{-\eta_1, \eta_2} + {i\over 4} g\zeta_{\eta _1, \eta _2}
\eea
Multiplying the first equation by $\s^3$, and combined the result with the second 
equation, we find the reduced dilatino equations,
\bea
(d) \qquad  0 = g\zeta_{\eta _1, \eta _2}
 = p_a \s^a \s^1 \zeta^*_{-\eta_1, \eta_2}
\eea

\subsection{The gravitino equation}

The gravitino equation is the same as equation (\ref{gravitino}).
The non-vanishing spin connection components are $\omega^{a} {}_{b}$ and 
\bea
\label{spincon-1}
\omega^m {}_n = \hat \omega^m {}_n
&\qquad&
\omega^m {}_{a} = e^m {D_{a} f_5 \over f_5}
\no\\
\omega^{i} {}_{j} = \hat \omega^{i} {}_{j}
&\qquad&
\omega^{i} {}_{a} = e^{i} {D_{a} f_3 \over f_3}
\eea
The hats refers to the canonical connections on $AdS_5$, $S^3$ respectively.
Projecting the spin-connection along the various directions we have the expression for $d\ep + \omega \ep$
\bea
(m) &\qquad& \nabla_m^\prime \ep + \half {D_{a} f_5 \over f_5} \G_{m} \G^{a} \ep
\no\\
(i) &\qquad& \nabla_{i}^\prime \ep + \half {D_{a} f_3 \over f_3} \G_{i} \G^{a} \ep
\no\\
(a) &\qquad& \nabla_{a} \ep
\eea
where the prime on the covariant derivative indicates that only the connection
along $AdS_5$, $S^3$ respectively is included.
Using the Killing spinor equations (\ref{KS-1}) we can eliminate the primed covariant derivatives,
which yields
\bea
(m) &\qquad& {1 \over 2 f_5} \G_{m}
\sum _{\eta _1, \eta _2}
\eta_1 \chi ^{\eta _1, \eta _2}
\otimes
\l( \zeta _{\eta _1, \eta _2}
\otimes u_+  + i \s^3 \zeta_{\eta _1, \eta _2} \otimes u_-  \r)
 + \half {D_{a} f_5 \over f_5} \G_{m} \G^{a} \ep
\no\\
(i) &\qquad& {i \over 2 f_3} \G_{i}
\sum _{\eta _1, \eta _2}
\eta_2 \chi ^{\eta _1, \eta _2}
\otimes
\l( -i \s^3 \zeta_{\eta _1, \eta _2} \otimes u_+
+ \zeta _{\eta _1, \eta _2}
\otimes u_-    \r)
+ \half {D_{a} f_3 \over f_3} \G_{i} \G^{a} \ep
\no\\
(a)  &\qquad& \nabla_{a} \ep
\eea
combining the terms, we have
\bea
(m) &\qquad&  \G_{m}
\sum_{\eta_1 ,\eta_2}  \chi^{\eta_1, \eta_2} \otimes
\left[ ({ \eta_1\over 2 f_5} \zeta_{\eta_1, \eta_2} - {1\over 2} {D_{a} f_5 \over f_5} \s^{a} \s^3 \zeta_{\eta_1, \eta_2} )
 \otimes u_+ \right.
\no \\
&&
\left. + ({ i \eta_1\over 2 f_5} \s^3 \zeta_{\eta_1, \eta_2} + {i\over 2} {D_{a} f_5 \over f_5} \s^{a} \zeta_{\eta_1, \eta_2} )
 \otimes u_-
\right ]
\no\\
(i) &\qquad&  \G_{i}
\sum_{\eta_1 ,\eta_2 }  \chi^{\eta_1, \eta_2} \otimes
\left[ ({ \eta_2\over 2 f_3} \s^3 \zeta_{\eta_1, \eta_2} - {1\over 2} {D_{a} f_3 \over f_3} \s^{a} \s^3\zeta_{\eta_1, \eta_2} )
 \otimes u_+ \right.
\no \\
&&
\left. + ({ i \eta_2\over 2 f_3} \zeta_{\eta_1, \eta_2} + {i\over 2} {D_{a} f_3 \over f_3} \s^{a} \zeta_{\eta_1, \eta_2} )
 \otimes u_-
\right ]
\no\\
(a)  &\qquad& \nabla_{a} \ep
\eea
where we have pulled a factor of $\G_M$ out front for equation (m), (i).  It will turn out that
all terms in the gravitino equation (m) (i)  will contain $\G_M \chi^{\eta_1, \eta_2 }$,
and we require the coefficients to vanish independently, just as we did for
the dilatino equation.  The coefficient of
$\G_M \chi^{\eta_1 ,\eta_2}, \chi^{\eta_3}$ can be expressed in the $\tau$-matrix
notation as
\bea
\label{redomega2-1}
(m) &\qquad& ({ 1 \over 2 f_5} \t^{30} - {1\over 2} {D_{a} f_5 \over f_5} \s^{a} \s^3 )\zeta
 \otimes u_+
 + ({ i \over 2 f_5} \t^{30} \s^3 + {i\over 2} {D_{a} f_5 \over f_5} \s^{a}) \zeta
 \otimes u_-
\no\\
(i) &\qquad& ({ 1\over 2 f_3} \t^{03}\s^3  - {1\over 2} {D_{a} f_3 \over f_3} \s^{a} \s^3 ) \zeta
 \otimes u_+
+ ({ i \over 2 f_3} \t^{03} + {i\over 2} {D_{a} f_3 \over f_3} \s^{a} )\zeta
 \otimes u_-
\eea
 The $F_{(5)}$ part is
\bea
{i \over 480} (\G \cdot F_{(5)}) \G^A e_A \ep
=  {i \over 2} f \G^{01234} \G^A e_A \ep
\eea
Projecting along the various directions, we have
\bea
(m) &\quad& \G_m \sum _{\eta _1, \eta _2}
\chi ^{\eta _1, \eta _2}
\otimes
\l( -{f\over 2} \zeta _{\eta _1, \eta _2}
\otimes u_+  - {if \over 2} \s^3 \zeta_{\eta _1, \eta _2} \otimes u_-  \r)
\no\\
(i) &\quad& \G_i  \sum _{\eta _1, \eta _2}
\chi ^{\eta _1, \eta _2}
\otimes
\l( {f\over 2} \zeta _{\eta _1, \eta _2}
\otimes u_+  + {if \over 2} \s^3 \zeta_{\eta _1, \eta _2} \otimes u_-  \r)
\no\\
(a) &\quad& \sum _{\eta _1, \eta _2}
\chi ^{\eta _1, \eta _2}
\otimes
\l( ({f \over 2} \s_a \s^3  - {i\over 2} q_a)\zeta_{\eta _1, \eta _2} \otimes u_+
+ ({i f\over 2} \s_a - {1\over 2}q_a \s^3)\zeta _{\eta _1, \eta _2}
\otimes u_-   \r)
\eea

\subsubsection{Assembling the complete gravitino BPS equation}

Now we combine the above equations to obtain the reduced gravitino equations.
We again argue that the $\Gamma_M \chi^{\eta_1, \eta_2}$ or $\chi^{\eta_1, \eta_2}$
are linearly independent which leads to the equations
\bea
(m) &~& 
0 = ({ 1\over 2 f_5} \t^{(30)}- {1\over 2} {D_{a} f_5 \over f_5} \s^{a} \s^3 - {f\over 2} )\zeta
 \otimes u_+
 + ({ i \over 2 f_5}\t^{(30)} \s^3 + {i\over 2} {D_{a} f_5 \over f_5} \s^{a} - {if\over 2}\s^3) \zeta
 \otimes u_-
\no\\
(i) &~& 
0 = ({ 1\over 2 f_3} \t^{(03)}\s^3 - {1\over 2} {D_{a} f_3 \over f_3} \s^{a} \s^3 + {f\over 2})\zeta
 \otimes u_+
+ ({ i \over 2 f_3}\t^{(03)} + {i\over 2} {D_{a} f_3 \over f_3} \s^{a} + {if \over 2} \s^3) \zeta
 \otimes u_-
\no \\
(a) &~&
0 =  (D_{a} + {i\over 2}   \hat \omega_{a} \s^3 + {f \over 2} \s_a \s^3 - {i\over 2} q_a ) \zeta
\otimes u_+
-i \s^3(D_{a} + {i\over 2} \hat \omega_{a} \s^3  +{ f\over 2} \s_a \s^3- {i\over 2} q_a )
\zeta \otimes u_- \no
\eea
Separating $u_+$ and $u_-$ terms reproduces the equations of (\ref{gravitino1-1}).

\section{Calculation of metric factors for $AdS_5 \times S^2 \times S^1 \times \Sigma$}
\label{metric}

We shall use combinations of the differential equations $(\pm)$
and of the algebraic gravitino BPS equations  of (\ref{BPS2a}) to bring out the corresponding relations. To this end, we compute,
\bea
D_z \left (\psi^{\dagger} \t^{(\a \b)}  \psi \right )
&=&
-  f  \psi^{\dagger} \t^{(\a \b)} \tau^{(12)}\xi^*
+ {i \over 4}  g_{\bar z}^* \xi^\dagger \t^{(\a \b)}  \psi
\no \\ &&
- {i \over 8} \left (  g_{z} \psi^{\dagger} \t^{(\a \b)}  \xi
+  g_{9}\psi^{\dagger} \t^{(\a \b)} \tau^{(13)} \psi^*
+ 3h \psi^{\dagger} \t^{(\a \b)}\t^{(12)}\psi^* \right )
\no \\
D_{z} \left ( \xi^{\dagger} \t^{(\a \b)}  \xi \right )
& = &
+  f \psi^{\dagger} \tau^{(12)} (\t^{(\a \b)})^t  \xi^*
+ {i \over 4}  g_{z}  \psi^{\dagger}\t^{(\a \b)}\xi
\no \\ &&
+ {i \over 8} \left ( - g_{\bar z}^*  \xi^{\dagger} \t^{(\a \b)}  \psi
+  g^*_{9}\xi^{\dagger} \t^{(\a \b)} \tau^{(13)} \xi^*
- 3h^*\xi^{\dagger} \t^{(\a \b)} \t^{(12)} \xi^* \right ) \qquad
\eea
In the subsequent subsections, we shall combine these results
with corresponding combinations from the algebraic gravitino BPS equations,
and seek relations of the following type,
\bea
\label{cons1a}
D_z \left ( r_1 \psi^{\dagger} \t^{(\a \b)}  \psi + r_2 \xi^{\dagger} \t^{(\a \b)}  \xi \right )
+
{D_z f_i \over f_i} \left ( r_3 \psi^{\dagger} \t^{(\a \b)}  \psi +
r_4 \xi^{\dagger} \t^{(\a \b)}  \xi \right ) =0
\eea
where $i=1,2,5$ and the coefficients $r_1, r_2,r_3,r_4$ may depend on $i$,
and $\alpha, \beta$, but  not on  $\Sigma$. We shall seek such relations
for {\sl generic fields $f_1,f_2, f_5, f, g_z, g_{\bar z}, h,$ and $g_9$}, though
we shall find that when special relations hold (such as $g_9=0$), additional
relations of type (\ref{cons1a}) exist.

\subsection{Relations involving $f_5$}

We start with the metric factor $f_5$, provided by the  $(m)$-equation.
Left-multiplying equation $(m_1)$  by $\psi^\dagger$, and left-multiplying
the complex conjugate of equation $(m_2)$ by $\xi ^\dagger$, we find,
\bea
{D_z f_5 \over f_5} \left ( \psi^{\dagger}   \t^{(\a \b)} \psi \right )
& = &
- {i \over f_5} \psi^{\dagger} \t^{(\a \b)} \t^{(22)} \xi^*
+  f \psi^{\dagger} \t^{(\a \b)}\t^{(12)} \xi^*
\no \\ &&
+{i \over 8} \left (  g_z \psi^{\dagger} \t^{(\a \b)}  \xi
-  g_{9} \psi^{\dagger} \t^{(\a \b)}  \tau ^{(13)} \psi^*
+ h \psi^{\dagger} \t^{(\a \b)} \t^{(12)} \psi^* \right )
\no \\
{D_z f_5 \over f_5} \left ( \xi^{\dagger} \t^{(\a \b)} \xi \right )
& = &
+ {i \over f_5} \xi^{\dagger} \t^{(\a \b)} \t^{(22)} \psi^*
+ f \xi^{\dagger} \t^{(\a \b)}\t^{(12)} \psi^*
\no \\ &&
+ {i \over 8} \left (  g^*_{\bar z} \xi^{\dagger} \t^{(\a \b)}  \psi
+   g^*_{9} \xi^{\dagger} \t^{(\a \b)} \tau ^{(13)} \xi^*
+  h^* \xi^{\dagger} \t^{(\a \b)} \t^{(12)} \xi^* \right )
\eea
For generic fields, term by term cancellation imposes the following requirements,
\bea
(f_5)  \qquad 0 & = &
- r_3 \t^{(\a \b)} \t^{(22)} + r_4 \tau ^{(22)}  ( \tau ^{(\a \b)}  )^t
\no \\
(f)  \qquad 0 & = &
(r_3- r_1) \t^{(\a \b)} \tau^{(12)} + (r_2-r_4) \tau^{(12)} (\t^{(\a \b)})^t
\no \\
(g_{\bar z}^*)  \qquad 0 & = &
2 r_1 - r_2 + r_4
\no \\
(g_z)  \qquad 0 & = &
2 r_2 - r_1+  r_3
\no \\
(g_9)  \qquad 0 & = &
( r_1  + r_3) \, g_9 \, \psi^{\dagger} \t^{(\a \b)} \tau^{(13)} \psi^*
=  ( r_2  + r_4) \, g_9^* \, \xi^{\dagger} \t^{(\a \b)} \tau^{(13)} \xi^*
\no \\
( h) \qquad 0 & = &
(3 r_1 - r_3) \, h \, \psi^{\dagger} \t^{(\a \b)}\t^{(12)}\psi^*
= (3 r_2 - r_4) \, h^* \, \xi^{\dagger} \t^{(\a \b)}\t^{(12)} \xi^*
\eea
We shall now systematically analyze and solve the above conditions.

\sm

If $r_3=0$, then equation $(f_5)$ implies $r_4=0$, and $(g_z)$ and $(g_{\bar z}^*)$
imply $r_1=r_2=0$, which yields the trivial solution.
Thus, we have $r_3 \not=0$, and by an overall rescaling, we choose $r_3=1$.
Equation $(f_5)$ then implies that we must have $|r_4|=1$.

\sm

Equations $(g_{\bar z} ^*)$ and $(g_z)$ then
reduce to $r_1 = 2r_2+1$, and $r_4=-3r_2-2$, supplemented with the condition
$|r_4|=|3r_2+2|=1$. These relations have two distinct solutions, according to
whether $r_4=\pm 1$. Satisfying the $(f)$ and $(f_5)$ equations then gives
two sets of solutions,
\bea
(r_1, r_2, r_3, r_4) = (-1,-1,1,1) \hskip 0.3in & \hskip 0.6in &
\tau ^{(\a \b)} \in \{ \tau^{(00)}, \tau^{(31)}, \tau^{(32)} , \t^{(33)} \}
\no \\
(r_1, r_2, r_3, r_4) = (1/3,-1/3,1,-1) & \hskip 0.6in &
\tau ^{(\a \b)} = \{ \tau^{(01)}, \tau^{(02)},\tau^{(03)}, \tau^{(30)}  \}
\eea
Equations $(h)$ automatically holds for both solutions,
while equations $(g_9)$ hold automatically
for the first solution, and with $\tau^{(\a \b)} \in \{ \tau ^{(01)}, \tau ^{(30)} \}$ for the second solution.
Thus, the first solution is {\sl generic}, and the second solution is generic
with $\tau^{(\a \b)} \in \{ \tau ^{(01)}, \tau ^{(30)} \}$.

\sm

\subsection{Relations involving $f_2$}

The analysis of the bilinear relations involving $f_2$ is analogous.
We start from the relations,
\bea
{D_z f_2 \over f_2} \left ( \psi^{\dagger}   \t^{(\a \b)} \psi \right )
& = &
- {i \over f_2} \psi^{\dagger} \t^{(\a \b)} \t^{(11)} \xi^*
-  f \psi^{\dagger} \t^{(\a \b)}\t^{(12)} \xi^*
\no \\ &&
+{i \over 8} \left ( - 3 g_z \psi^{\dagger} \t^{(\a \b)} \xi
+ 3 g_{9} \psi^{\dagger} \t^{(\a \b)}  \tau ^{(13)} \psi^*
+ h \psi^{\dagger} \t^{(\a \b)} \t^{(12)} \psi^* \right )
\no \\
{D_z f_2 \over f_2} \left ( \xi^{\dagger} \t^{(\a \b)} \xi \right )
& = &
- {i \over f_2} \xi^{\dagger} \t^{(\a \b)} \t^{(11)} \psi^*
-  f \xi^{\dagger} \t^{(\a \b)}\t^{(12)} \psi^*
\no \\ &&
- {i \over 8} \left (  3 g^*_{\bar z} \xi^{\dagger} \t^{(\a \b)}  \psi
+3   g^*_{9} \xi^{\dagger} \t^{(\a \b)} \tau ^{(13)} \xi^*
-  h^* \xi^{\dagger} \t^{(\a \b)} \t^{(12)} \xi^* \right )
\eea
For generic fields, term by term cancellation imposes the following requirements,
\bea
(f_2)  \qquad 0 & = &
 r_3 \t^{(\a \b)} \t^{(11)} + r_4 \tau ^{(11)}  ( \tau ^{(\a \b)}  )^t
\no \\
(f)  \qquad 0 & = &
- (r_1+ r_3) \t^{(\a \b)} \tau^{(12)} + (r_2 + r_4) \tau^{(12)} (\t^{(\a \b)})^t
\no \\
(g_{\bar z}^*)  \qquad 0 & = &
2 r_1 - r_2 - 3 r_4
\no \\
(g_z)  \qquad 0 & = &
2 r_2  - r_1 - 3 r_3
\no \\
(g_9)  \qquad 0 & = &
( 3 r_3  - r_1) \, g_9 \, \psi^{\dagger} \t^{(\a \b)} \tau^{(13)} \psi^*
= ( 3 r_4 - r_2 ) \, g_9 ^* \, \xi^{\dagger} \t^{(\a \b)} \tau^{(13)} \xi^*
\no \\
( h) \qquad 0 & = &
(3 r_1 - r_3) \, h \, \psi^{\dagger} \t^{(\a \b)}\t^{(12)}\psi^*
= (3 r_2 - r_4) \, h^* \, \xi^{\dagger} \t^{(\a \b)}\t^{(12)}\xi^*
\eea
We may again set $r_3=1$, so that $|r_4|=1$ without loss of generality.
Equations $(g_{\bar z} ^*)$ and $(g_z)$ reduce to $r_1 = 2r_2-3$, and
$r_4=r_2-2$, supplemented with the condition
$|r_4|=|r_2-2|=1$. Equations $(f)$ and $(f_2)$ leave
two distinct solutions, given by
\bea
(r_1, r_2, r_3, r_4) = (-1,1,1,-1) & \hskip 0.6in &
\tau ^{(\a \b)} \in \{ \tau^{(00)}, \tau^{(01)},\tau^{(02)} ,\tau^{(10)}, \tau^{(11)} ,
\no \\
& \hskip 0.6in & \hskip 0.6in \tau^{(12)}, \tau^{(20)},\tau^{(21)},  \tau^{(22)} , \tau^{(33)} \}
\no \\
(r_1, r_2, r_3, r_4) = (3,3,1,1) \hskip 0.2in & \hskip 0.6in &
\tau ^{(\a \b)} \in \{  \tau^{(31)} , \tau^{(32)} \}
\eea
Equations $(h), (g_9)$ automatically hold for the second solution.
They hold for the first solution with $\tau ^{(\a \b)} =  \tau^{(33)}$, which is generic;
with $\tau^{(00)} , \tau^{(10)},\tau^{(20)} $ when $g_9=0$,
with $\tau^{(01)}, \tau ^{(11)},\tau^{(21)}$ when $h=0$, and
with $\tau^{(02)}, \tau ^{(12)}, \tau^{(22)}$ when $g_9=h=0$.

\subsection{Relations involving $f_1$}

The analysis of the bilinear relations involving $f_1$ is analogous.
We start from the relations,
\bea
{D_z f_1 \over f_1} \left ( \psi^{\dagger}   \t^{(\a \b)} \psi \right )
& = &
{ \nu \over f_1} \psi^{\dagger} \t^{(\a \b)} \t^{(23)} \xi^*
-  f \psi^{\dagger} \t^{(\a \b)}\t^{(12)} \xi^*
\no \\ &&
+{i \over 8} \left (  g_z \psi^{\dagger} \t^{(\a \b)} \xi
+ 3 g_{9} \psi^{\dagger} \t^{(\a \b)}  \tau ^{(13)} \psi^*
-3 h \psi^{\dagger} \t^{(\a \b)} \t^{(12)} \psi^* \right )
\no \\
{D_z f_1 \over f_1} \left ( \xi^{\dagger} \t^{(\a \b)} \xi \right )
& = &
- { \nu \over f_1} \xi^{\dagger} \t^{(\a \b)} \t^{(23)} \psi^*
- f \xi^{\dagger} \t^{(\a \b)}\t^{(12)} \psi^*
\no \\ &&
+ {i \over 8} \left (   g^*_{\bar z} \xi^{\dagger} \t^{(\a \b)}  \psi
- 3   g^*_{9} \xi^{\dagger} \t^{(\a \b)} \tau ^{(13)} \xi^*
- 3 h^* \xi^{\dagger} \t^{(\a \b)} \t^{(12)} \xi^* \right ) \qquad
\eea
For generic fields, term by term cancellation imposes the following requirements,
\bea
(f_1)  \qquad 0 & = &
 r_3 \t^{(\a \b)} \t^{(23)} + r_4 \tau ^{(23)}  ( \tau ^{(\a \b)}  )^t
\no \\
(f)  \qquad 0 & = &
- (r_1+ r_3) \t^{(\a \b)} \tau^{(12)} + (r_2 + r_4) \tau^{(12)} (\t^{(\a \b)})^t
\no \\
(g_{\bar z}^*)  \qquad 0 & = &
2 r_1 - r_2 + r_4
\no \\
(g_z)  \qquad 0 & = &
2 r_2  - r_1 + r_3
\no \\
(g_9)  \qquad 0 & = &
( 3 r_3  - r_1) \, g_9 \, \psi^{\dagger} \t^{(\a \b)} \tau^{(13)} \psi^*
= ( 3 r_4 - r_2 ) \, g_9 ^* \, \xi^{\dagger} \t^{(\a \b)} \tau^{(13)} \xi^*
\no \\
( h) \qquad 0 & = &
(r_1 + r_3) \, h \, \psi^{\dagger} \t^{(\a \b)}\t^{(12)}\psi^*
= (r_2 + r_4) \, h^* \, \xi^{\dagger} \t^{(\a \b)}\t^{(12)}\xi^*
\eea
We may again set $r_3=1$, so that $|r_4|=1$ without loss of generality.
Equations $(g_{\bar z} ^*)$ and $(g_z)$ reduce to $r_1 = 2r_2+1$, and
$r_4=-3r_2-2$, supplemented with the condition
$|r_4|=|3r_2+2|=1$. Equations $(f)$ and $(f_1^{-1})$ leave
two distinct solutions, given by
\bea
(r_1, r_2, r_3, r_4) = (-1,-1,1,1) \hskip 0.2in & \hskip 0.5in &
\tau ^{(\a \b)} \in \{ \tau^{(01)}, \tau^{(10)}, \tau^{(12)} ,\tau^{(13)}, \tau^{(20)},
\no \\
& \hskip 0.6in & \hskip 0.6in  \tau^{(22)},\tau^{(23)},\tau^{(30)}, \tau^{(32)} , \tau^{(33)}  \}
\no \\
(r_1, r_2, r_3, r_4) = (1/3,-1/3,1,-1) &  &
\tau ^{(\a \b)} \in \{  \tau^{(02)} , \tau^{(03)}, \tau^{(11)} , \tau^{(21)} \}
\eea
Equations $(h)$ automatically hold for the first solution, while $(g_9)$
automatically holds only with $\tau ^{(\a \b)}\in \{ \tau ^{(01)}, \tau ^{(30)},
\tau^{(32)},   \tau ^{(33)} \}$, which is the generic solution.
It holds with $\tau^{(10)}, \tau^{(12)}$, $\tau^{(13)}$, $\tau^{(20)}, \tau^{(22)},\tau^{(23)} $
when $g_9=0$. For the second solution, the equations hold with
$ \tau ^{(11)},\tau^{(21)}$ when $h=0$, and with $\tau^{(02)}, \tau ^{(03)}$
when $g_9=h=0$.

\section{Vanishing Hermitian forms  for $AdS_5 \times S^2 \times S^1 \times \Sigma$}
\label{appVHF}

In this Appendix, we derive three sets of vanishing Hermitian forms.

\subsection{First set of Hermitan relations}

The first set of Hermitian relations is obtained by considering the combination
of the BPS equations of (\ref{BPS2a}), given by $2(m) + (i) + (9)$. 
All $f, g_z , g_{\bar z}, h, h^*$ terms cancel in these combinations, 
and the resulting equations are,
\bea
& &
D_z \ln (f_5^2 f_2 f_1) \psi + \left({2i \over f_5} \t^{(22)} + {i\over f_2} \t^{(11)}
- {\nu \over f_1} \t^{(23)}\right) \xi^* - {i\over 2} g_9 \t^{(13)} \psi^* = 0
\no \\
&&
D_z \ln (f_5^2 f_2 f_1) \xi + \left(-{2i \over f_5} \t^{(22)} + {i\over f_2} \t^{(11)}
+ {\nu \over f_1} \t^{(23)}\right) \psi^* + {i\over 2} g^*_9 \t^{(13)} \xi^* = 0
\eea
Multiplying the first equation by $\xi^t \t^{(\a \b)}$, the second by $-\psi^t \t^{(\a \b)t}$,
adding both to cancel the differential terms, and taking the transpose, we obtain,
\bea
0& =&  \xi^\dagger  \left({2i \over f_5} \t^{(22)} + {i\over f_2} \t^{(11)}
+ {\nu \over f_1} \t^{(23)}\right) {\t^{(\a \b)}} ^t \xi
- {i\over 2} g_9 \psi^\dagger \t^{(13)} {\t^{(\a \b)}}^t  \xi
\no \\
&&
+\psi^\dagger  \left( {2i \over f_5} \t^{(22)} - {i\over f_2} \t^{(11)}
+ {\nu \over f_1} \t^{(23)}\right) \t^{(\a \b)} \psi
- {i\over 2} g^*_9 \xi^\dagger \t^{(13)}  \t^{(\a \b)} \psi
\eea
Each term in $\xi ^\dagger \tau \xi$ or $\psi ^\dagger \tau \psi$, and each combination
of terms involving $g_9$ and $g_9^*$, is either real or purely imaginary.
Defining the sign factors $\pm$ associated with each label $(\a \b)$,
\bea
\t^{(\a \b)*} = \pm \t^{(13)} \t^{(\a \b)} \t^{(13)}
\eea
the corresponding term in $g_9$ and $g_9^*$ is real for the $-$ sign  and
imaginary for the $+$ sign. Separating out the corresponding real and imaginary
parts yields valuable information.
We summarize the resulting relations below, the first column corresponding
to the $(\a \b)$-assignment(s) from which the equation has been derived.
\bea
\label{vh1}
(00) & \hskip 0.5in &
H_+^{(23)} =0
\hskip 0.5in
{2 \over f_5} H_+^{(22)}
- {1 \over f_2} H_-^{(11)}
- \half H_{g+}^{(13)} =0
\no \\
(01) &&
H_{g+}^{(12)} =0
\hskip 0.5in
 {1\over f_2} H_-^{(10)}
- {\nu \over f_1} H_+^{(22)}=0
\no \\
(02) &&
H_+^{(13)} =0
\hskip 0.5in
{2 \over f_5} H_-^{(20)}
- {\nu \over f_1} H_-^{(21)}
- { i \over 2} H_{g-}^{(11)} =0
\no \\
(03) &&
H_{g+}^{(10)} =0
\hskip 0.5in
{2\over f_5} H_+^{(21)}
+{1\over f_2} H_-^{(12)}
-{\nu \over f_1} H_+^{(20)}=0
\no \\
(10) &&
H_+^{(32)} =0
\hskip 0.5in
 {1\over f_2} H_-^{(01)}
+ {\nu \over f_1} H_+^{(33)}
+  {1\over 2} H_{g+}^{(03)} = 0
\no \\
(11) &&
H_{g+}^{(02)} =0
\hskip 0.5in
{2\over f_5} H_+^{(33)}
+ {1 \over f_2} H_-^{(00)}=0
\no \\
(12) &&
H_{g-}^{(01)} =0
\hskip 0.5in
{2\over f_5} H_-^{(30)}
+{1\over f_2} H_+^{(03)}
-{\nu \over f_1} H_-^{(31)}=0
\no \\
(13) &&
H_-^{(02)}  =0
\hskip 0.5in
{2 \over f_5} H_+^{(31)}
- {\nu \over f_1} H_+^{(30)}
- \half H_{g+}^{(00)} =0
\no \\
(20) &&
H_{g-}^{(33)} =0
\hskip 0.5in
{1\over f_2} H_+^{(31)}
+ {\nu \over f_1} H_-^{(03)}=0
\no \\
(21) &&
H_-^{(02)}  =0
\hskip 0.5in
{2 \over f_5} H_-^{(03)}
+ {1 \over f_2} H_+^{(30)}
-{i \over 2}  H_{g-}^{(32)} =0
\no \\
(22) && \hskip 1.15in
{2 \over f_5} H_+^{(00)}
+ {1 \over f_2} H_-^{(33)}
- {\nu \over f_1} H_+^{(01)}
- \half  H_{g+}^{(31)} =0
\no \\
(23) &&
H_{g-}^{(30)} =0
\hskip 0.5in
{2\over f_5} H_-^{(01)}
- {\nu \over f_1} H_-^{(00)}=0
\no \\
(30) &&
H_+^{(13)} =0
\hskip 0.5in
{2 \over f_5} H_+^{(12)}
+ {1 \over f_2} H_-^{(21)}
+ \half  H_{g+}^{(23)} =0
\no \\
(31) &&
H_{g+}^{(22)} =0
\hskip 0.5in
{1\over f_2} H_-^{(20)}
+ {\nu \over f_1} H_+^{(12)}=0
\no \\
(32) &&
H_+^{(23)} =0
\hskip 0.5in
{2 \over f_5} H_-^{(10)}
- {\nu \over f_1} H_-^{(11)}
+ {i \over 2}  H_{g-}^{(21)} =0
\no \\
(33) &&
H_{g+}^{(20)} =0
\hskip 0.5in
{2\over f_5} H_+^{(11)}
- {1\over f_2} H_-^{(22)}
-{\nu \over f_1} H_+^{(10)}=0
\eea

\subsection{Second set of Hermitian relations}

The second set of Hermitian equalities is derived from eliminating
$D_z f_5$, $D_z f_2$, $D_z f_1$, $g_z$, and $g_{\bar z }^*$ terms in each set
of equations (m), (i), or (9) respectively. Because the (9) equation can be obtained from a
linear superpostion of (m) and (i) equations, and those from the first set, we only need to
calculate the (m) and (i) equations.

\subsubsection{The $f_5$-equation}

For the $f_5$ equation, we multiply the first by $\xi^t \t^{(\a \b)}$, and the second by
$\psi^t \t^{(\a \b)}$. The $g_z, g_{\bar z}$ terms
will disappear if $\t^{(\a \b)}$ is anti-symmetric, $\t^{(\a \b)t} = -\t^{(\a \b)}$.
Adding both equations to eliminate the $D_z f_5$-term, we obtain,
\bea
 0& =& \xi^t \t^{(\a \b)}\left( {i \over  f_5} \tau^{(22)} \xi^* - f \tau^{(12)} \xi^*
 + {i \over 8}  g_9 \tau ^{(13)} \psi ^* - {i \over 8}h \tau^{(12)} \psi^* \right)
\no\\
&& + \psi^t \t^{(\a \b)}\left( -{i \over  f_5} \tau^{(22)} \psi^* - f \tau^{(12)}    \psi^*
 - {i \over 8}   g^*_9 \tau ^{(13)} \xi^* - {i \over 8} h^* \tau^{(12)} \xi^*\right)
\eea
Separating the real and imaginary parts for all antisymmetric $\t^{(\a\b)}$, we obtain,
\bea
\label{vh2}
(02) & \hskip 0.5in &
H_+^{(10)}  =0
\hskip 1.1in
{i \over f_5} H_-^{(20)}
- {1 \over 8} H_{g-}^{(11)}
- {i \over 8} H_{h+}^{(10)}=0
\no \\
(20) &&
H_{h+}^{(32)}=0
\hskip 1.1in
{i \over f_5} H_-^{(02)}
- i f  H_{+}^{(32)}
+ {1 \over 8} H_{g-}^{(33)}=0
\no \\
(12) &&
H_{g-} ^{(01)} + i H_{h+}^{(00)} =0
\hskip 0.5in
{1\over f_5} H_-^{(30)}
 -  f H_+^{(00)}  = 0
\no \\
(21) &&
H_{h+}^{(33)}=0
\hskip 1.1in
{1\over f_5} H_-^{(03)}
- f H_+^{(33)}
+ {i\over 8} H_{g-}^{(32)}=0
\no \\
(32) &&
H_{+}^{(20)}=0
\hskip 1.1in
{1 \over f_5} H_-^{(10)}
- {i \over 8} H_{g-}^{(21)}
+ {1 \over 8} H_{h+}^{(20)}=0
\no \\
(23) &&
H_{g-} ^{(30)} + i H_{h+}^{(31)}=0
\hskip 0.5in
{1\over f_5} H_-^{(01)}
 - f H_+^{(31)} = 0
\eea

\subsubsection{The $f_2$-equation}

For the $f_2$ equations, with a similar procedure, we have the following result,
\bea
&& \xi^t \t^{(\a \b)}\left( {i \over  f_2} \tau^{(11)} \xi^* + f \tau^{(12)} \xi^*
 - {3i \over 8}  g_9 \tau ^{(13)} \psi ^* - {i \over 8}h \tau^{(12)} \psi^* \right)
\no\\
&& + \psi^t \t^{(\a \b)}\left( {i \over  f_2} \tau^{(11)} \psi^* + f \tau^{(12)}\psi^*
 + {3i \over 8}   g^*_9 \tau ^{(13)} \xi^* - {i \over 8} h^* \tau^{(12)} \xi^*\right) = 0
\eea
Separating into real and imaginary parts, we obtain,
\bea
\label{vh3}
(02) & \hskip 0.5in &
{1\over f_2} H_+^{(13)} +  f H_+^{(10)} = 0
\hskip 0.5in
3 H_{g-}^{(11)} -i H_{h+}^{(10)}=0
\no \\
(20) &&
i f H_+^{(32)} -  {3\over 8}  H_{g-}^{(33)} = 0
\hskip 0.5in
{1 \over f_2} H_+ ^{(31)} +{1 \over 8} H^{(32)} _{h+} =0
\no \\
(12) &&
{1\over f_2} H_+^{(03)} +  f H_+^{(00)} = 0
\hskip 0.5in
3 H_{g-}^{(01)} -i H_{h+}^{(00)}=0
\no \\
(21)
&&
H_{h+}^{(33)} =0
\hskip 1.3in
{1\over f_2} H_+^{(30)} + f H_+^{(33)} - {3i\over 8} H_{g-}^{(32)}=0
\no \\
(32)
&&
3i H_{g-}^{(21)}+  H_{h+}^{(20)} =0
\hskip 0.6in
{1\over f_2} H_+^{(23)} + f H_+^{(20)} = 0
\no \\
(23) &&
H_+ ^{(31)} =0
\hskip 1.3in
{1\over f_2} H_+^{(32)}  - {3i\over 8} H_{g-}^{(30)} - { 1 \over 8} H_{h+} ^{(31)} =0
\quad
\eea
Similar equations may be derived for the (9)-equations. In view of the first
set of relations however, these will all be linearly dependent on the equations
already obtained,and will thus not be needed.

\subsection{Third set of Hermitian relations}

A third set of Hermitian relations is obtained by eliminating $g_z$ and $g_{\bar z}^*$
terms between equations (m), (i), and (9), contracting the resulting equations
with symmetric $\t^{(\a\b)}$, taking differences to eliminate $D_z f$ term, and 
forming the linear combinations $ (m)-(9)$,
\bea
D_{z} \ln (f_5/ f_1)    \psi
+\left ({i \over  f_5} \tau^{(22)} +{\nu\over  f_1}  \tau^{(23)} \right )\xi^*
 -  2 f \tau^{(12)} \xi^*
 + {i \over 2} (   g_9 \tau ^{(13)} \psi ^* - h \tau^{(12)} \psi^*) = 0
\no\\
D_{ z} \ln (f_5 /f_1)   \xi
- \left ({i \over  f_5} \tau^{(22)} + { \nu \over  f_1}  \tau^{(23)} \right )\psi^*
- 2f \tau^{(12)}    \psi^*
 - {i \over 2} (  g^*_9 \tau ^{(13)} \xi^* +  h^* \tau^{(12)} \xi^*) = 0
\eea
Multiplying the above equations by $\xi^t \t^{(\a\b)}$ and $\psi^t \t^{(\a\b)}$ respectively,
with $\t^{(\a\b)t} = \t^{(\a\b)}$, and taking the difference to eliminate the $D_z \ln (f_5/f_1)$
terms, we obtain,
\bea
&&
\xi^t \t^{(\a\b)}\left({i \over  f_5} \tau^{(22)}\xi^* +{\nu\over  f_1}  \tau^{(23)}\xi^*
 -  2 f \tau^{(12)} \xi^*
 + {i \over 2} g_9 \tau ^{(13)} \psi ^* - {i \over 2} h \tau^{(12)} \psi^* \right)
\no\\ &&
+ \psi^t \t^{(\a\b)}\left({i \over  f_5} \tau^{(22)}\psi^* +{ \nu \over  f_1}  \tau^{(23)}\psi^*
+ 2f \tau^{(12)}    \psi^*
 + {i \over 2} g^*_9 \tau ^{(13)} \xi^* + {i \over 2}  h^* \tau^{(12)} \xi^*\right) = 0
  \qquad
\eea
Separating real and imaginary parts, we have
\bea
\label{vh4}
(00) & \hskip 0.5in &
{1 \over  f_5}H_+^{(22)} + {1 \over 2} H_{g+}^{(13)} = 0
\hskip 0.5in
-{\nu\over  f_1} H_{+}^{(23)} -2 f H_-^{(12)} + {i \over 2} H_{h-}^{(12)}=0
\no \\
(01) &&
{1 \over  f_5}H_+^{(23)} - {1 \over 2} H_{g+}^{(12)} = 0
\hskip 0.5in
{i\nu\over  f_1} H_{+}^{(22)} - 2i f H_{-}^{(13)} - {1 \over 2} H_{h-}^{(13)}=0
\no \\
(03)&&
 {1 \over  f_5}H_+^{(21)} + {\nu\over  f_1} H_{+}^{(20)}=0
\hskip 0.45in
2i f H_{-}^{(11)}
- {i \over 2} H_{g+}^{(10)}
+ {1 \over 2} H_{h-}^{(11)}=0
\no\\
(10) &&
{\nu\over  f_1} H_{+}^{(33)}
+ {1 \over 2} H_{g+}^{(03)} =0
\hskip 0.5in
{1 \over  f_5}H_+^{(32)}
- 2 f H_{-}^{(02)}
+ {i \over 2} H_{h-}^{(02)}=0
\no \\
(11) &&
{\nu\over  f_1} H_{+}^{(32)}
+ {1 \over 2} H_{g+}^{(02)} =0
\hskip 0.5in
{i \over  f_5}H_+^{(33)}
- 2 i f H_{-}^{(03)}
- {1 \over 2} H_{h-}^{(03)}=0
\no \\
(13) && \qquad
{i \over  f_5} H_+^{(31)}
+ {i\nu\over  f_1} H_{+}^{(30)}
- 2i f H_{-}^{(01)}
+{i \over 2}  H_{g+}^{(00)}
- {1 \over 2} H_{h-}^{(01)}=0
\no \\
(22) &&\qquad
{i\over  f_5} H_{+}^{(00)}
+{i\nu \over  f_1}H_{+}^{(01)}
- 2if H_{-}^{(30)}
+{i \over 2}  H_{g+}^{(31)}
- {1 \over 2} H_{h-}^{(30)}=0
\no \\
(30) &&
{1\over  f_5} H_{+}^{(12)}
- {1 \over 2} H_{g+}^{(23)}= 0
\hskip 0.5in
-{i\nu \over  f_1}H_{+}^{(13)}
+ 2 if H_{-}^{(22)}
+ {1 \over 2} H_{h-}^{(22)}=0
\no \\
(31) &&
 {1\over  f_5} H_{+}^{(13)}
+ {1 \over 2} H_{g+}^{(22)}= 0
\hskip 0.5in
{\nu \over  f_1}H_{+}^{(12)}
+ 2 f H_{-}^{(23)}
- {i \over 2} H_{h-}^{(23)}=0
\no \\
(33)
&& {1\over  f_5} H_{+}^{(11)}
+ {\nu \over  f_1}H_{+}^{(10)}=0
\hskip 0.5in
- 2 f  H_{-}^{(21)}
+ {1 \over 2} H_{g+}^{(20)}
+ {i \over 2} H_{h-}^{(21)}=0
\eea

\end{document}